\documentclass[preprint]{aastex63}

\usepackage{xspace}
\usepackage{graphicx}
\usepackage{natbib}
\usepackage{url}
\usepackage{bm}
\usepackage{amsmath}
\usepackage[maxfloats=256]{morefloats}

\newcommand{\arcsper}{\rlap.{^{\prime\prime}}}
\newcommand{\yr}{\ensuremath{\,{\rm yr}}}

\newcommand{\gyr}{\ensuremath{\,{\rm Gyr}}}
\newcommand{\khz}{\ensuremath{\,{\rm kHz}}}
\newcommand{\mhz}{\ensuremath{\,{\rm MHz}}}
\newcommand{\ghz}{\ensuremath{\,{\rm GHz}}}

\newcommand{\K}{\ensuremath{\,{\rm K}}}

\newcommand{\m}{\ensuremath{\,{\rm m}}}

\newcommand{\percc}{\ensuremath{\,{\rm cm^{-3}}}}

\newcommand{\kpc}{\ensuremath{\,{\rm kpc}}}

\newcommand{\kms}{\ensuremath{\,{\rm km\, s^{-1}}}}
\newcommand{\msun}{\ensuremath{\,M_\odot}}

\newcommand{\hr}{\,hr}

\newcommand{\jy}{\,Jy}
\newcommand{\mjy}{\,mJy}

\newcommand{\mjyb}{\ensuremath{\rm \,mJy\,beam^{-1}}}

\newcommand{\microjyb}{\ensuremath{\rm \,\mu Jy\,beam^{-1}}}

\newcommand{\hna}{\ensuremath{\langle {\rm Hn}\alpha \rangle}}
\newcommand{\hnb}{\ensuremath{\langle {\rm Hn}\beta \rangle}}
\newcommand{\hng}{\ensuremath{\langle {\rm Hn}\gamma \rangle}}

\newcommand{\expo}[1]{\ensuremath{10^{#1}}}
\newcommand{\nexpo}[2]{\ensuremath{#1 \times 10^{#2}}}

\newcommand{\hei}{He\,{\sc i}}
\newcommand{\hii}{H\,{\sc ii}}

\newcommand{\ciii}{C\,{\sc iii}}

\newcommand{\he}[1]{\ensuremath{^{#1}{\rm He}}}
\newcommand{\hep}[1]{\ensuremath{^{#1}{\rm He}^{+}}}

\newcommand{\li}[1]{\ensuremath{{}^#1{\rm Li}}}

\newcommand{\her}[1]{\ensuremath{{}^#1{\rm He}/{\rm H}}}
\newcommand{\hepr}[1]{\ensuremath{{}^#1{\rm He}^{+}/{\rm H^+}}}
\newcommand{\heppr}[1]{\ensuremath{{}^#1{\rm He}^{++}/{\rm H^+}}}
\newcommand{\cratio}{\ensuremath{{}^{12}{\rm C}/^{13}{\rm C}}}
\newcommand{\threec}[1]{3C\thinspace #1}
\newcommand{\ic}[1]{IC\thinspace #1}
\newcommand{\ngc}[1]{NGC\thinspace #1}

\newcommand\urltilda{\kern -.15em\lower .7ex\hbox{\~{}}\kern .04em}
\newcommand{\gsim}{\ensuremath{\gtrsim}}

\begin{document}

\title{Do All Low-Mass Stars Undergo Extra Mixing Processes?}

\author[0000-0002-2465-7803]{Dana S. Balser}
\affiliation{National Radio Astronomy Observatory, 520 Edgemont Rd.,
  Charlottesville, VA 22903, USA}

\author[0000-0003-0640-7787]{Trey V. Wenger}
\affiliation{Dominion Radio Astrophysical Observatory, Herzberg
  Astronomy and Astrophysics Research Centre, National Research
  Council, P.O. Box 248, Penticton, BC V2A 6J9, Canada}

\author[0000-0003-4866-460X]{T. M. Bania} \affiliation{Institute
  for Astrophysical Research, Astronomy Department, Boston University,
  725 Commonwealth Avenue, Boston, MA 02215, USA}

\begin{abstract}

  Standard stellar evolution models that only consider convection as a
  physical process to mix material inside of stars predict the
  production of significant amounts of \he3\ in low-mass stars
  ($M < 2$\msun), with peak abundances of
  $^{3}{\rm He/H} \sim \nexpo{\rm few}{-3}$ by number.  Over the
  life-time of the Galaxy, this ought to produce \her3\ abundances
  that diminish with increasing Galactocentric radius.  Observations
  of \hep3\ in \hii\ regions throughout the Galactic disk, however,
  reveal very little variation in the \he3\ abundance with values of
  \her3\ similar to the primoridal abundance,
  $(^{3}{\rm He/H})_{\rm p} \sim \expo{-5}$.  This discrepancy, known
  as the ``\he3\ Problem'', can be resolved by invoking in stellar
  evolution models an extra-mixing mechanism due to the thermohaline
  instability.  Here, we observe \hep3\ in the planetary nebula J320
  (PN G190.3--17.7) with the Jansky Very Large Array (JVLA) to confirm
  a previous \hep3\ detection made with the VLA that supports standard
  stellar yields.  This measurement alone indicates that not all stars
  undergo extra mixing.  Our more sensitive observations do not detect
  \hep3\ emission from J320 with an RMS noise of 58.8\microjyb\ after
  smoothing the data to a velocity resolution of 11.4\kms.  We
  estimate an abundance limit of $^{3}{\rm He/H} \le \nexpo{2.75}{-3}$
  by number using the numerical radiative transfer code NEBULA.  This
  result nullifies the last significant detection of \hep3\ in a PN
  and allows for the possibility that all stars undergo extra mixing
  processes.

\end{abstract}

\keywords{Stellar evolution - Planetary nebulae - Radio lines}

\section{Background}\label{sec:background}

The \he3\ isotope is one of the few elements that is not only produced
several minutes after the Big Bang during the era of primordial
nucleosynthesis but also is subsequently made inside stars via stellar
nucleosynthesis \citep[e.g.,][]{boesgaard85}.  Measurements of \he3\
therefore provide a unique probe of cosmic evolution.  \citet{rood76}
first identified the significance of measuring the \he3\ abundance in
the interstellar medium (ISM).  They predicted an enrichment of the
primordial \he3\ abundance due to stellar nucleosynthesis based on
\he3\ yields from low-mass stars ($M < 2$\msun).  \citet{rood76}
argued that (1) the present day ISM \her3\ abundance ratio should be
significantly larger than the protosolar value; (2) \her3\ should grow
with source metallicity; and (3) there should be a radial gradient in
\her3\ abundance across the Milky Way disk with higher abundances in
the more processed central regions.

Detection of \he3\ has proven challenging, however, since there is
expected to be about one \he3\ atom for every 10,000 \he4\ atoms.
Isotopic shifts for light elements are small compared with typical
line widths, so using He recombination lines (e.g., \hei\
$\lambda\,6678$) to detect \he3\ is difficult.  Nevertheless,
anomalously high \her3\ abundance ratios have been detected with He
recombination lines in some stars \citep[e.g.,][]{sargent61}.  These
very high \he3\ abundances are thought to be due to diffusion and are
therefore not representative of typical abundances. There have also
been anomalously high \her3\ abundances detected from in situ
measurements of Solar energetic particle events
\citep[e.g.,][]{wiedenbeck20}.  Potential molecular transitions
including \he3\ are rare since helium is inert and seldom found in
molecular form.  Detection of HeH$^{+}$ in the PN \ngc{7027} is a
recent exception \citep{gusten19}.

\citet{townes57} was the first to suggest the \hep3\ hyperfine
transition at 8665.650\mhz\ \citep{novick58} as a possible
astrophysical tracer at radio frequencies.  \citet{goldwire67}
calculated the Einstein coefficient of the \hep3\ hyperfine
transition, corresponding to a relatively short radiative lifetime of
16,000\yr, indicating the plausibility of measuring \he3\ in \hii\
regions \citep[also see][]{syunyaev66}.  Initial attempts at detecting
\hep3\ in \hii\ regions were unsuccessful and limited by high receiver
system temperatures \citep{seling69, predmore71}.

\citet{rood79} made the first detection of \he3\ in the ISM toward the
giant \hii\ region W51 with the Max-Planck Institut f\"{u}r
Radioastronomie (MPIfR) 100\m\ telescope .  They derived a \her3\
abundance ratio similar to the protosolar value and thus found no
evidence for the production of \he3\ in low-mass stars.  Observations
of \hep3\ in \hii\ regions over the last four decades have yielded
similar results---stars are not significant producers of \he3\
\citep{rood84, bania87, balser94, bania97, balser18}.

Accurate determination of the \her3\ abundance ratio, the
astrophysical quantity of interest, requires models of the density and
ionization structure of the \hii\ region.  This is because the tracer
of \he3, the hyperfine transition, is sensitive to
$\int n_{\rm e} \, d\ell$, whereas the tracer of H, the free-free
continuum, is sensitive to $\int n_{\rm e}^{2} \, d\ell$
\citep{balser99a}.  Here $n_{\rm e}$ is the electron density and
$d\ell$ is the path length across the \hii\ region.  Ionization
structure is important because H and He have different ionization
potentials \citep{bania07}.  Using this information, \citet{bania02}
selected sources with simple morphologies that would produce the most
accurate \her3\ abundance ratios and found that \her3\ was
approximately constant across the Galactic disk---``The \he3\
Plateau.''  They suggested that the \he3\ Plateau abundance of
$^{3}{\rm He/H} = 1.1 \pm 0.2 \times 10^{-5}$ by number is the
primordial abundance.  This was later confirmed by combining results
from the Wilkinson Microwave Anisotropy Probe (WMAP) with Big Bang
Nucleosynthesis (BBN) models yielding a primordial abundance of
$(^{3}{\rm He/H})_{\rm p} = 1.00 \pm 0.07 \times 10^{-5}$ \citep{romano03,
  cyburt08}.

Since low-mass stars were expected to be sources of \he3\ enrichment
to the ISM via mass loss during the asymptotic giant branch (AGB)
phase, similar efforts were made to detect \hep3\ in
PNe. \citet{rood92} made the first detection of \hep3\ in the PN
\ngc{3242} with the MPIfR 100\m\ \citep[also see][]{balser97,
  balser99b}.  They derived an abundance of
$^{3}{\rm He/H} \gsim 10^{-3}$, two orders of magnitude larger than
abundances found in \hii\ regions, consistent with standard stellar
models.  Observations of \hep3\ were made for a handful of PNe over
the next two decades \citep{balser97, balser99b, balser06b,
  guzman-ramirez13, guzman-ramirez16, bania21}.  \hep3\ detections
were also claimed in PNe J320 \citep{balser06b} and \ic{418}
\citep{guzman-ramirez16} with derived abundance ratios of
$^{3}{\rm He/H} \sim {\rm few} \times 10^{-3}$.  So there seemed to be
solid evidence that some low-mass stars were producing copious amounts
of \he3\ to be returned to the ISM during the PN phase.

Galactic chemical evolution (GCE) models using \he3\ yields from
standard stellar evolution predict significantly larger \he3\
abundances over the lifetime of the Milky Way than are observed in
\hii\ regions \citep{galli95, galli97, olive95}.  Most GCE models also
predict negative radial \her3\ abundance ratio gradients within the
Galactic disk because the central regions have undergone more stellar
processing than the outer over the lifetime of the Milky Way.  This is
inconsistent with the \he3\ Plateau revealed by observations.
Moreover, in situ measurements of helium within the Jovian atmosphere
with the Galileo Probe yield
$^{3}{\rm He}/^{4}{\rm He} = (1.66 \pm 0.05) \times 10^{-4}$
\citep{mahaffy98}.  This corresponds to a protosolar abundance of
$^{3}{\rm He/H} = (1.5 \pm 0.2) \times 10^{-5}$, indicating very
little production of \he3\ over the past 4.5\gyr.  \citet{galli97}
called these discrepancies ``The \he3\ Problem.''

\citet{rood84} suggested that some sort of mixing could be taking
place in low-mass stars that might explain the lower observed \he3\
abundances than expected in the ISM.  They posited that such a mixing
mechanism may be related to the destruction of \li7\ in main-sequence
stars and low \cratio\ abundance ratios observed in low-mass red giant
branch (RGB) stars \citep[also see][]{charbonnel95, hogan95, weiss96}.
Numerous studies indicate that some sort of extra mixing is occurring
when low-mass stars reach the luminosity bump on the RGB
\citep[e.g.,][]{gilroy89, luck94, charbonnel98b, gratton00,
  pilachowski03, smiljanic09}.  The luminosity bump occurs when the
hydrogen burning shell reaches the chemical discontinuity created by
the maximum extent of the convective envelope during the first
dredge-up.  For many years rotation-induced mixing was thought to be
the main mechanism responsible for the abundance anomalies
\citep[e.g.,][]{sweigart79, charbonnel95, charbonnel98b, boothroyd99},
but more accurate stellar evolution simulations that treat the
transport of angular momentum by meridional circulation and shear
turbulence self consistently do not produce enough mixing around the
luminosity bump to account for the observed surface abundance
variations \citep{palacios06}.

A breakthrough occurred when \citet{eggleton06} constructed
three-dimensional stellar evolution models and discovered the
destabilizing role played by the molecular weight inversion that is
produced at the external edge of the hydrogen-burning shell by the
\he3(\he3, 2p)\he4\ reaction.  \citet{charbonnel07a} pointed out that
the first instability to occur under these conditions is a
double-diffusive instability called the thermohaline instability
\citep{stern60}.  As the molecular weight gradient increases, the
temperature has a stabilizing effect since the timescale for thermal
diffusion is shorter than the time it takes for the material to mix.

Stellar evolution models that incorporate thermohaline mixing are able
to account for the anomalous \cratio\ and \li7\ abundances that are
observed in low-mass stars and predict \he3\ yields that are
significantly reduced compared to standard models
\citep{charbonnel07a, denissenkov08, eggleton08, charbonnel10,
  cantiello10, lagarde11}. \citet{lagarde12b} used these stellar
yields together with GCE models to predict a modest enrichment of
\he3\ with time that is consistent with \hii\ region observations in
the Milky Way disk \citep[also see][]{balser18}.

There are some outstanding issues that remain concerning the treatment
of the thermohaline instability just after the luminosity bump on the
RGB \citep[e.g., see][]{karakas14}.  The diffusion coefficient is
proportional to $C$, a dimensionless free parameter which is related
to the aspect ratio of the ``salt'' fingers.  \citet{charbonnel07a}
use values of $C \sim 1000$ because experiments favor thin fingers
instead of blobs \citep{ulrich72}.  Numerical simulations of
thermohaline convection, however, predict a lower value for $C$ than
is necessary to solve the \he3\ Problem
\citep[e.g.,][]{denissenkov11}.  Rotation may also influence the
effectiveness of thermohaline mixing \citep[e.g.,][]{maeder13,
  sengupta18}.

The fact that a few PNe have estimated \her3\ abundances consistent
with standard \he3\ yields implies that the thermohaline instability
is not effective in all low-mass stars.  \citet{charbonnel98a}
estimate that 4\% of red giant stars have \cratio\ abundance ratios
that are consistent with expectations from standard stellar models.
GCE models that allow 4\% of low-mass stars to produce standard \he3\
yields \citep{lagarde12b} are still consistent with \hii\ region
observations \citep{balser18}.  \citet{eggleton08} suggested that deep
mixing of \he3\ and CNO isotopes is not optional and that this
mechanism would destroy most of the \he3\ produced on the main
sequence.  \citet{charbonnel07b} proposed that fossil magnetic fields
in red giant stars that are descendants of Ap stars could inhibit
thermohaline mixing.  In sum, stellar modeling has yet to reach a
theoretical consensus concerning the fate of \he3\ produced by stellar
nucleosynthesis.

There is some question, however, whether \hep3\ has been detected in
any PNe.  Using the Green Bank Telescope (GBT), \citet{bania21} have
recently shown that the reported \hep3\ detection in \ngc{3242} with
the MPIfR 100\m, and confirmed with independent observations made with
the NRAO 140 Foot telescope \citep{balser99b}, is not real.  This
incorrect result probably stems from systematic errors due to standing
waves caused by reflections from the telescope superstructure.
Observations of \hep3\ from PNe are very challenging since the low
ionized mass produces very weak \hep3\ intensities that are at the
limits of most radio facilities.  The clear aperture of the GBT
reduced these systematic errors in the spectral baselines by an order
of magnitude.  Because of these systematic effects in traditionally
designed radio telescopes, \citet{bania21} were skeptical of the
claimed detection of \hep3\ in \ic{418} with the NASA Deep Space
Station 63 (DSS-63) telescope \citep{guzman-ramirez16}.  The lack of
any serious tests of the spectral baselines together with
discrepancies in the radio recombination line (RRL) parameters make
this claimed detection dubious.  The only remaining detection of
\hep3\ in a PN that seems plausible is for J320 observed with the VLA
\citep{balser06b}.  Interferometers have an advantage over single-dish
telescopes in that many instrumental spectral baseline effects are
removed because the signals between two antennas are correlated.  Here
we discuss new JVLA observations for the PN J320 made to confirm our
previous \hep3\ VLA detection.

\section{Observations and Data Reduction}\label{sec:obs}

Interferometers typically have stable spectral baselines but they are
not perfect.  The \citet{balser06b} J320 VLA observations suffered
from three problems: (1) a 3.3\mhz\ ripple common to all antennas
caused by reflections within the waveguide; (2) a limited number of
spectral channels and bandwidth which together provided very few
channels for characterizing the spectral baselines; and (3) only one
RRL transition available to assess the accuracy of these measurements.
The latter two problems were due to limitations with the VLA
correlator.  \citet{bania21} have shown that tuning to many RRLs
simultaneously can be used to assess the accuracy of the spectral
baselines and constrain models of the nebula to derive accurate \her3\
abundance ratios.

The JVLA overcomes all three of these problems with the VLA
observations.  Optical fiber has replaced the old waveguides and the
3.3\mhz\ ripple is gone.  The JVLA Wideband Interferometric Digital
ARchitecture (WIDAR) correlator provides us with an ample number of
channels across a large bandwidth to accurately measure the spectral
baseline.  The flexibility of WIDAR allows us to tune to many RRLs
simultaneously to carefully assess the quality of the spectral
baselines.  For example, adjacent RRLs should have similar line
profiles and we know the intensity ratios of various RRLs in local
thermodynamic equilibrium (LTE).  Moreover, these RRLs, together with
the free-free continuum can be used to constrain the nebular model
required to derive accurate \her3\ abundance ratios.

We therefore used the JVLA at X-band (8--10\ghz) in the
D-configuration to observe \hep3\ in the PN J320 to confirm the VLA
\hep3\ detection.  Hereafter, we distinguish between the two J320
\hep3\ data sets using the project codes: VLA (AB0794) and JVLA
(21A-005).  Table~\ref{tab:jvla} summarizes the observations.  We
observed for a total time of 29\hr\ to achieve a similar
signal-to-noise ratio (S/N) as our previous VLA observations.  The
half-power beam-width (HPBW) of the primary beam (field of view) is
about 5\arcmin\ at the \hep3\ frequency of 8665.650\mhz, and in the
D-configuration the synthesized HPBW is about 10\arcsec.

We configured the WIDAR correlator so that each of the two 1\ghz\
basebands was tuned to 8, 128\mhz\ wide spectral windows at full
polarization to observe the free-free continuum emission and 12,
16\mhz\ wide spectral windows at dual polarization to observe various
spectral lines.  The 128\mhz\ ``continuum'' windows each had 64
channels corresponding to a spectral resolution of 2.00\mhz\ and
covering a total of 2\ghz.  The 16\mhz\ ``spectral line'' windows each
had 512 channels corresponding to a spectral resolution of 31.25\khz\
($\sim 1$\kms\ at 9\ghz).  We sampled the \hep3\ hyperfine transition
in two spectral windows for redundancy together with these RRL
transitions: 7 Hn$\alpha$, 7 Hn$\beta$, and 8 Hn$\gamma$ (see
Table~\ref{tab:spws} for details).  Here n is the principal quantum
number and $\alpha$, $\beta$, $\gamma$ correspond to
$\Delta{\rm n} = 1$, 2, 3.  The 16\mhz\ bandwidth provides a velocity
span of $\sim 500$\kms, sufficient to include the corresponding
Hen$\alpha$, Hen$\beta$, and Hen$\gamma$ transitions.  The H113$\beta$
RRL is blended with the H129$\gamma$ RRL and therefore these
transitions were not observed.

\begin{deluxetable}{lc}
\tabletypesize{\small}
\tablecaption{JVLA Observational Summary \label{tab:jvla}}
\tablehead{
\colhead{Parameters} & \colhead{J320} 
}
\startdata
Project\dotfill                           & 21A-005 \\
Dates\dotfill                             & 2021 April 23--2021 May 25 \\
Total Time (hr)\tablenotemark{\scriptsize a}\dotfill  & 29 \\
Configuration\dotfill                     & D \\
R.A. of field center (J2000)\dotfill      & 05:05:34.56 \\
Decl. of field centere (J2000)\dotfill    & 10:42:26.60 \\
LSR central velocity (\kms)\dotfill       & $-37.9$ \\
Primary beam FWHM (arcmin)\dotfill        & $\sim 5$ \\
Synthesized beam FWHM (arcsec)\dotfill    & $\sim 10$ \\
Continuum bandwidth (GHz)\dotfill         & 2 \\
Line bandwidth (MHz)\dotfill              & 16 \\
Number of spectral channels\dotfill       & 512 \\
Spectral resolution (\khz)\dotfill        & 31.25 \\
Velocity resolution (\kms)\dotfill        & $\sim 1$ \\
Velocity span (\kms)\dotfill              & $\sim 500$ \\
Flux density/bandpass calibrator\dotfill  & J0542+4951 (\threec{147}) \\
Gain calibrator\dotfill                   & J0530+1331 \\
Continuum rms (\microjyb)\dotfill         &  25 (bandwidth: 16\mhz) \\
Line channel rms (\microjyb)\dotfill      &  135 (channel width: 2.5\kms)\\ 
\enddata
\tablenotetext{\scriptsize a}{Wall clock time that includes calibration, slew time, etc.}
\end{deluxetable}

\begin{deluxetable}{llccc}
\tabletypesize{\scriptsize}
\tablecaption{JVLA Spectral Windows \label{tab:spws}}
\tablehead{
  \colhead{Spectral} & \colhead{Center Freq.}\tablenotemark{\scriptsize a} & \colhead{}
  & \colhead{Bandwidth} & \colhead{} \\
  \colhead{Window} & \colhead{(MHz)} & \colhead{Transition}
  & \colhead{(MHz)} & \colhead{Channels}
}
\startdata
\cutinhead{Continuum}
1  & 8057.50009 & \dots & 128 & 64 \\
2  & 8185.50009 & \dots & 128 & 64 \\
3  & 8313.50009 & \dots & 128 & 64 \\
4  & 8441.50009 & \dots & 128 & 64 \\
5  & 8569.50009 & \dots & 128 & 64 \\
6  & 8697.50009 & \dots & 128 & 64 \\
7  & 8825.50009 & \dots & 128 & 64 \\
8  & 8953.50009 & \dots & 128 & 64 \\
9  & 9092.52938 & \dots & 128 & 64 \\
10 & 9220.52938 & \dots & 128 & 64 \\
11 & 9348.52938 & \dots & 128 & 64 \\
12 & 9476.52938 & \dots & 128 & 64 \\
13 & 9604.52938 & \dots & 128 & 64 \\
14 & 9732.52938 & \dots & 128 & 64 \\
15 & 9860.52938 & \dots & 128 & 64 \\ 
16 & 9988.52938 & \dots & 128 & 64 \\
\cutinhead{Spectral Line}
17 & 8665.65  & \hep3\ & 16 & 512 \\
18 & 8665.65  & \hep3\ & 16 & 512 \\
19 & 8045.605 & H93$\alpha$ & 16 & 512 \\
20 & 8309.385 & H92$\alpha$ & 16 & 512 \\
21 & 8584.823 & H91$\alpha$ & 16 & 512 \\
22 & 8872.571 & H90$\alpha$ & 16 & 512 \\
23 & 9173.324 & H89$\alpha$ & 16 & 512 \\
24 & 9487.824 & H88$\alpha$ & 16 & 512 \\
25 & 9816.867 & H87$\alpha$ & 16 & 512 \\
26 & 8213.052 & H116$\beta$ & 16 & 512 \\
27 & 8427.316 & H115$\beta$ & 16 & 512 \\
28 & 8649.099 & H114$\beta$ & 16 & 512 \\
29 & 9116.569 & H112$\beta$ & 16 & 512 \\
30 & 9362.976 & H111$\beta$ & 16 & 512 \\
31 & 9618.343 & H110$\beta$ & 16 & 512 \\
32 & 9883.083 & H109$\beta$ & 16 & 512 \\
33 & 8293.843 & H132$\gamma$ & 16 & 512 \\
34 & 8483.082 & H131$\gamma$ & 16 & 512 \\
35 & 8678.122 & H130$\gamma$ & 16 & 512 \\
36 & 9086.512 & H128$\gamma$ & 16 & 512 \\
37 & 9300.343 & H127$\gamma$ & 16 & 512 \\
38 & 9520.936 & H126$\gamma$ & 16 & 512 \\
39 & 9748.561 & H125$\gamma$ & 16 & 512 \\
40 & 9983.501 & H124$\gamma$ & 16 & 512 \\
\enddata
\tablecomments{The continuum windows span a total of $\sim 2$\ghz.
  The spectral line windows have a velocity span of $\sim 500$\kms\
  and a spectral resolution of $\sim 1$\kms.}
\tablenotetext{\scriptsize a}{Rest frequencies are listed for spectral
  line windows.}
\end{deluxetable}

We observed J320 between 2021 April 23 and 2021 May 25 during seven
distinct epochs, each with a duration of 4--5\hr.  We started each
epoch by observing the flux density calibrator J0542+4951
(\threec{147}), which was also used to set the delays and calibrate
the bandpass.  We then observed our PN J320 interleaved with
observations of the gain calibrator J0530+1331 every $\sim 20$
minutes.

We use the Wenger Interferometry Software Package (WISP) to calibrate
and image our JVLA data \citep{wenger18}.  WISP is a Python wrapper
for the Common Astronomy Software Package \citep[CASA;][]{mcmullin07}.
Here we follow the calibration and imaging procedures discussed in
Appendix A of \citet{wenger19a}.  The calibration procedures consist
of flagging bad data, calculating the calibration solutions, and
applying the calibration solutions.  This is an iterative process that
includes both automatic and manual flagging.

WISP automatically generates clean images from the calibrated
visibility data.  For the free-free continuum and \hep3\ spectral line
windows we use the native synthesized HPBW of
$9\arcsper9 \times 9\arcsper2$ when deconvolving the beam from the
dirty image.  In contrast, for the RRLs we first smooth the images to
a common spatial resolution of 12\arcsec\ since we want to average
(stack) RRLs with the same order (e.g., Hn$\alpha$,
$\Delta {\rm n} = 1$).  These RRLs have different frequencies and
therefore different synthesized HPBWs.  We denote these stacked
spectra as \hna, \hnb, and \hng.  Spectra within the data cubes are
smoothed and regridded to a common velocity resolution of 2.5\kms.
Since typical line widths in PNe are $\gsim 30$\kms, we also generate
data cubes with a coarse velocity resolution\footnote{We specifically
  chose a velocity resolution of 11.4\kms\ to be consistent with the
  resolution used by \citet{balser06b} for their reported \hep3\
  detection in J320.} of 11.4\kms.  The CASA task {\tt TCLEAN}
generates the following images and data cubes: (1) a multi-scale,
multi-frequency synthesis (MS-MFS) continuum image by combining all 16
continuum windows; (2) an MS-MFS image of each continuum and spectral
line window; and (3) a multi-scale data cube of each spectral line
window. Unless noted the continuum is not subtracted from the spectral
line data products.

\section{Results}\label{sec:results}

\begin{figure}
  \centering
  \includegraphics[angle=0,scale=1.0]{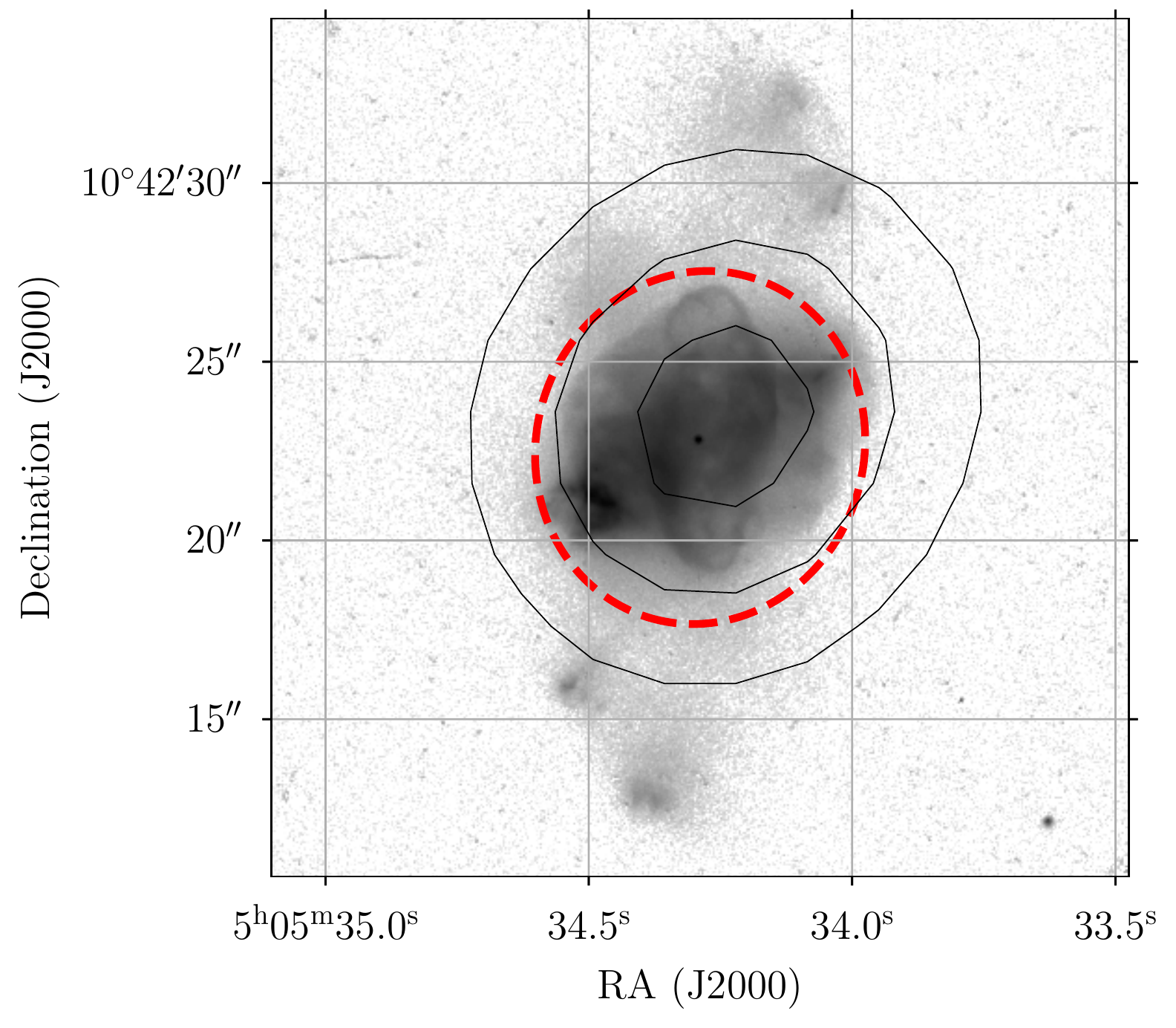}
  \caption{HST H$\alpha$ and JVLA radio continuum image of J320.  {\it
      Greyscale:} HST H$\alpha$ emission of J320 \citep[see][original
    observing program: Borkowski PI 6347]{harman04}.  {\it Contours:}
    JVLA MS-MFS continuum emission of J320 in the \hep3\ spectral
    window with contour levels at 5, 10, and 15\mjyb.  The center
    frequency is 8665.650\mhz\ and the bandwidth is 16\mhz.  The RMS
    noise in the image is 25\microjyb.  The synthesized HPBW of
    $9\arcsper9 \times 9\arcsper2$ is represented by the dashed red
    ellipse.}
  \label{fig:cont}
\end{figure}

The PN J320, discovered by \citet{jonckheere16}, is both spatially and
kinematically complex.  The object has two or three bipolar lobes
surrounded by high speed knots together with a surrounding halo
\citep{harman04, rechy-garcia20}.  The JVLA MS-MFS radio continuum
emission of J320 together with its H$\alpha$ emission image is shown
in Figure~\ref{fig:cont}.  J320 is just resolved by the JVLA,
consistent with the Hubble Space Telescope (HST) H$\alpha$ size of
about 10\arcsec\ \citep{harman04}.  The complex nature of this PN is
therefore not visible in our radio data.  The RMS noise in the image
is 25\microjyb, sufficient to detect the free-free continuum emission
from J320 with a ${\rm S/N} > 500$.  The integrated continuum flux
density is 23\mjy\ at the \hep3\ frequency of 8665.65\mhz, consistent
with previous VLA results \citep[cf.,][]{balser06b}.

\begin{figure}
  \centering
  \includegraphics[angle=0,scale=0.45]{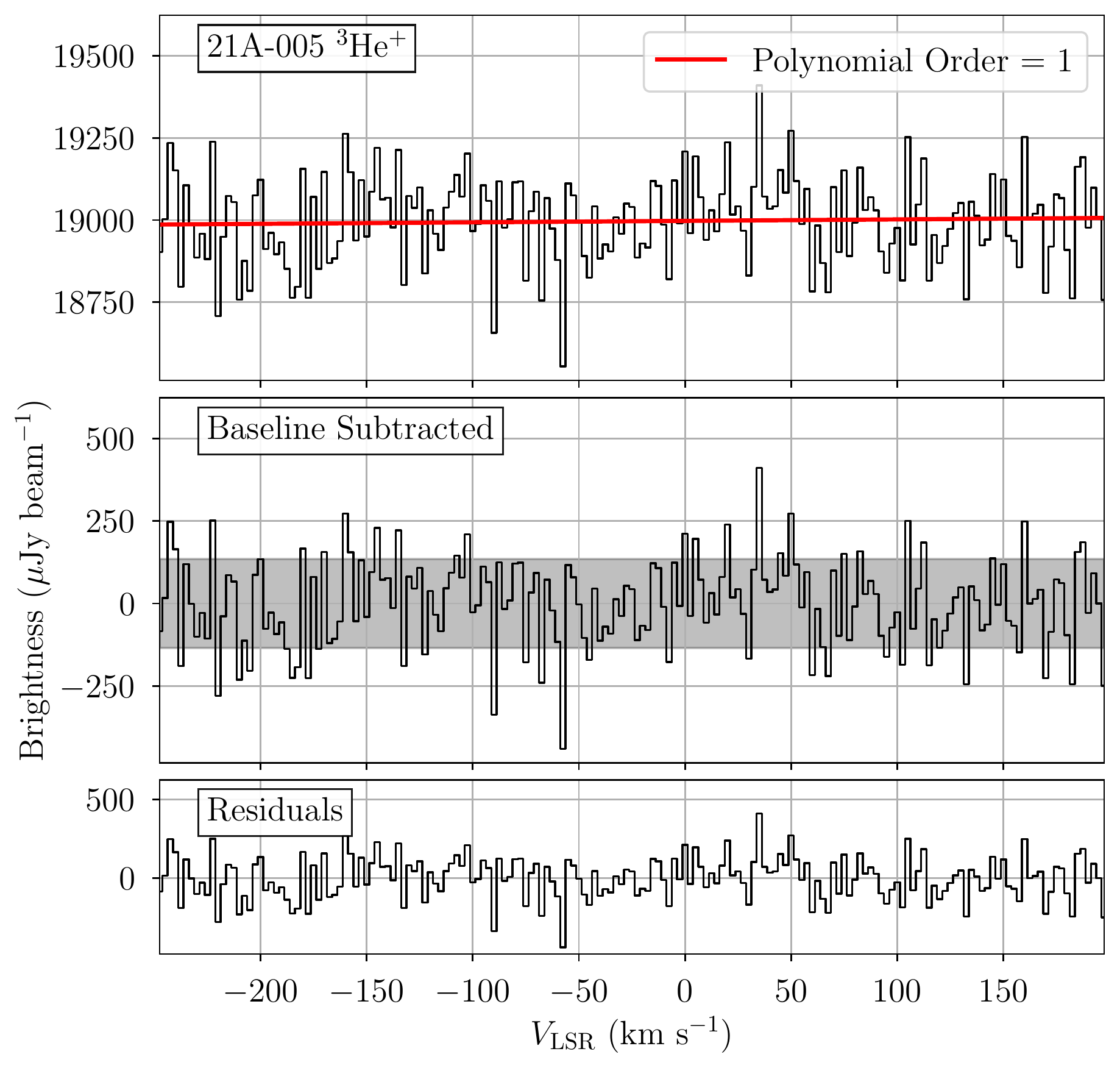}
  \includegraphics[angle=0,scale=0.45]{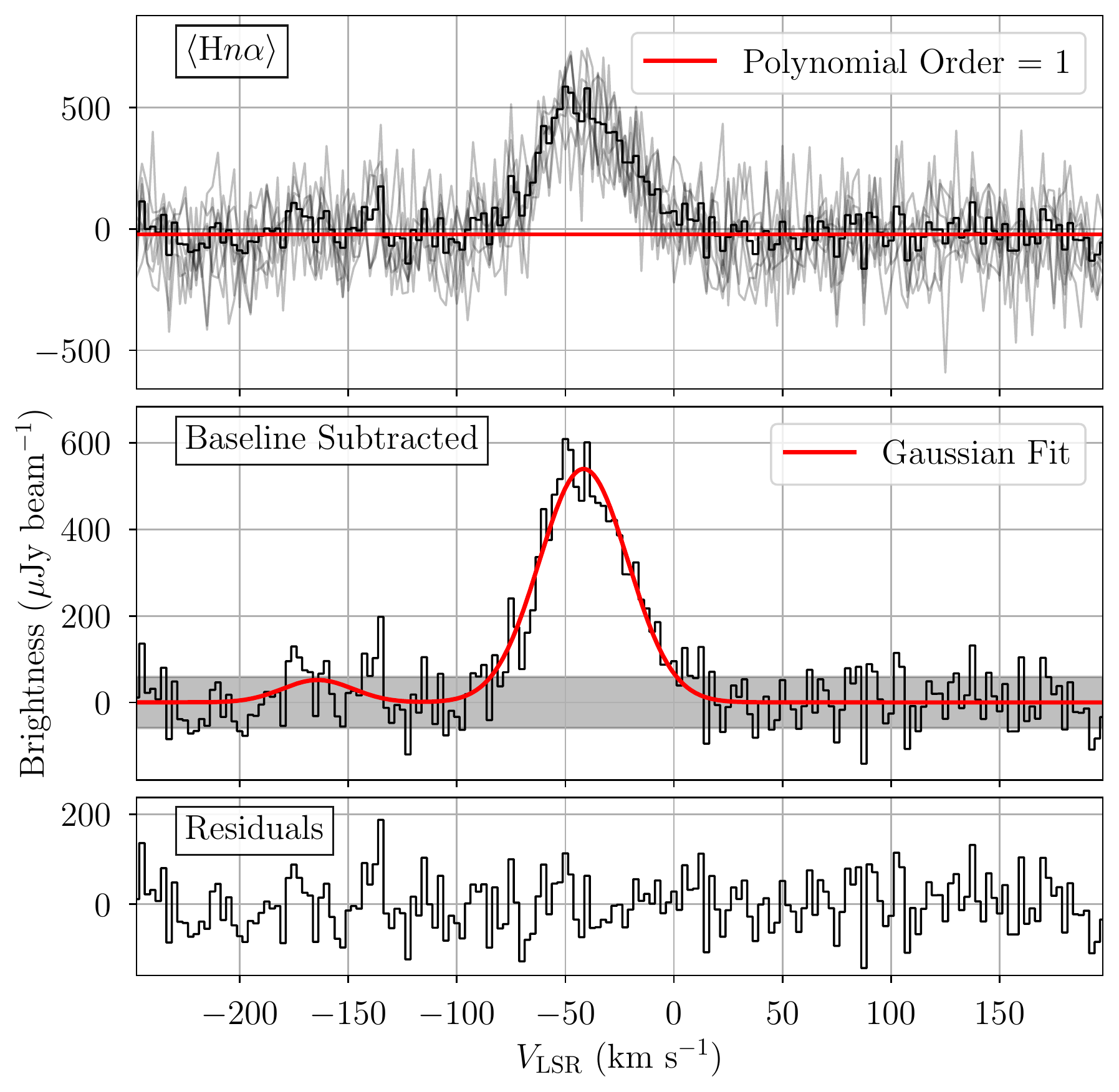}
  \includegraphics[angle=0,scale=0.45]{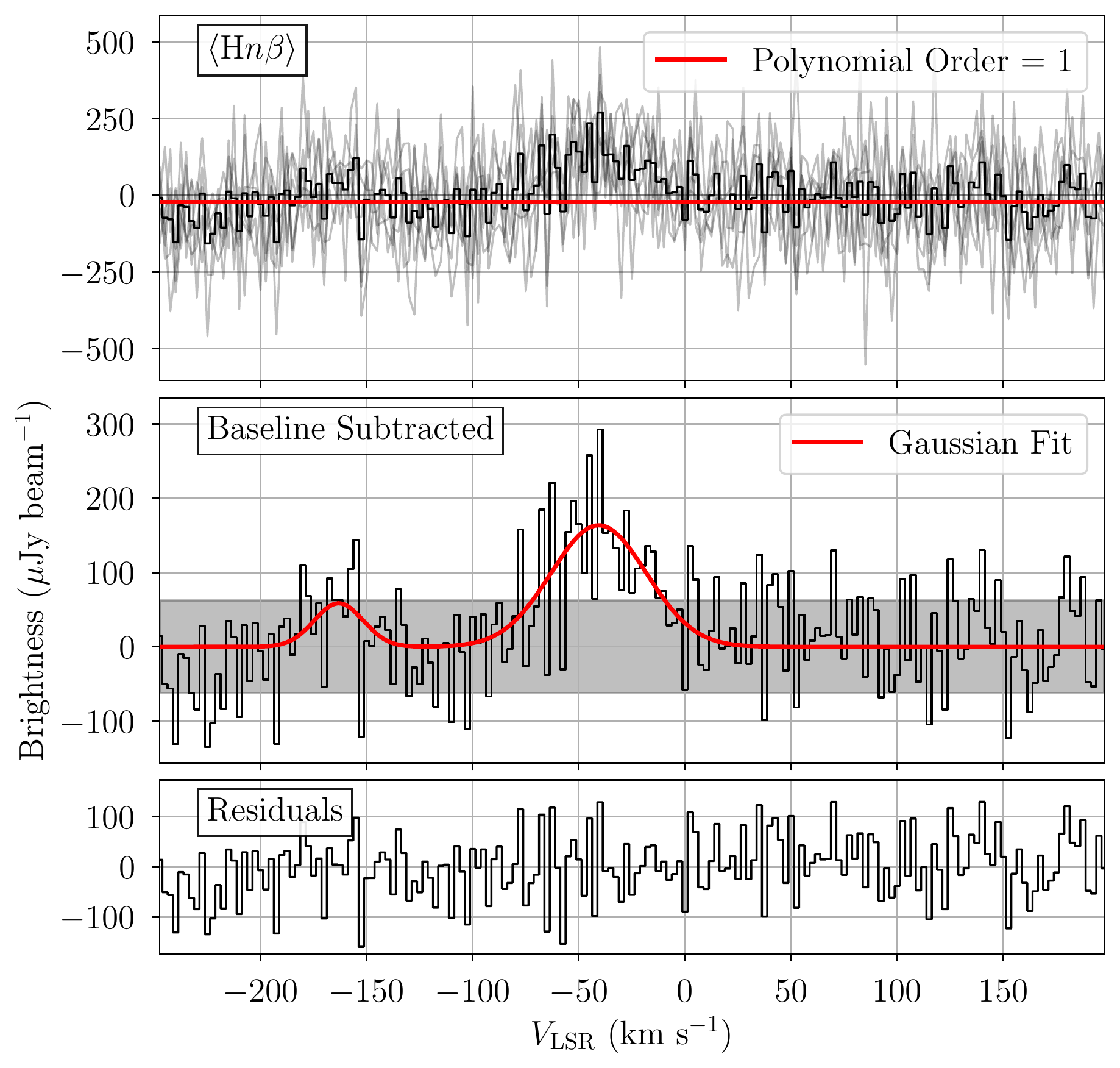}
  \includegraphics[angle=0,scale=0.45]{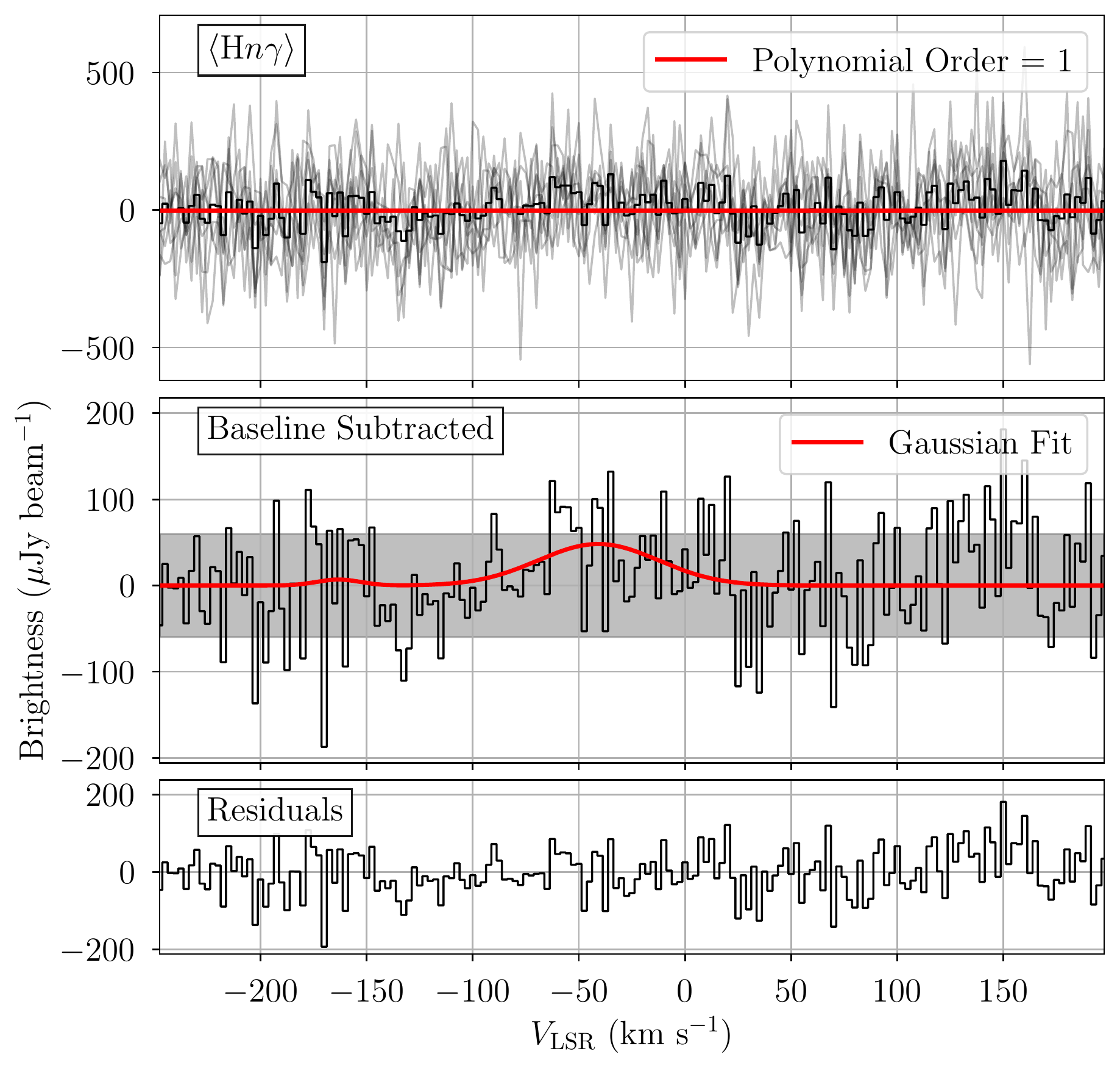}
  \caption{JVLA spectra of J320: \hep3\ (top-left), \hna\ (top-right),
    \hnb\ (bottom-left), and \hng\ (bottom-right).  Spectra are
    extracted from the brightest pixel in the continuum image and are
    displayed as histograms.  The spectra are smoothed and regridded
    to a velocity resolution of 2.5\kms.  {\it Top panel:} Spectrum
    with the first-order polynomial baseline model (red curve).  For
    the RRL plots the light gray curves are individual RRL spectra and
    the black curve corresponds to the stacked RRL spectrum.  {\it
      Middle panel:} Baseline-subtracted spectrum with a two-component
    Gaussian fit (red curve) for the RRL spectra only.  Here we fit
    the baseline model and the Gaussian components simultaneously. The
    He RRL component is assumed to be shifted $-122.47$\kms\ with
    respect to the H component.  The shaded region represents the 68\%
    confidence interval ($\pm 1\,\sigma$) spectral RMS measured in the
    residual spectrum.  {\it Bottom panel:} Residuals of the Gaussian
    fit subtracted from the data.}
\label{fig:spec}
\end{figure}

Figure~\ref{fig:spec} shows \hep3\ and stacked RRL spectra for the
spectral pixel (spaxel) in the data cube that corresponds to the
brightest region in the continuum image.  There is no clear visual
evidence of a \hep3\ line, but we do detect \hii\ RRL emission in the
\hna\ and \hnb\ spectra and perhaps in the \hng\ spectrum. We
therefore fit Gaussian profiles to the H and He RRLs shown by the red
curves in the middle panels of Figure~\ref{fig:spec}.  Specifically,
we simultaneously fit a first-order polynomial and two Gaussian
profiles to the entire spectral window.  We fix the location of the He
component with respect to the H component by $-122.47$\kms; that is,
we do not fit for the center velocity of the He component but rather
assume the shift produced by the mass of the heavier He nucleus.

Detecting \hep3\ emission in PNe is challenging.  Instrumental spectral
baseline structure can mimic or mask the wide, weak \hep3\ spectral
transition.  We therefore need to rigorously assess the quality of the
data and, in particular, the robustness of the non-detection of \hep3\
in J320.  We employ several statistical techniques to demonstrate that
the \hep3\ spectrum is consistent with random Gaussian noise.

\begin{figure}
  \centering
  \includegraphics[angle=0,scale=0.60]{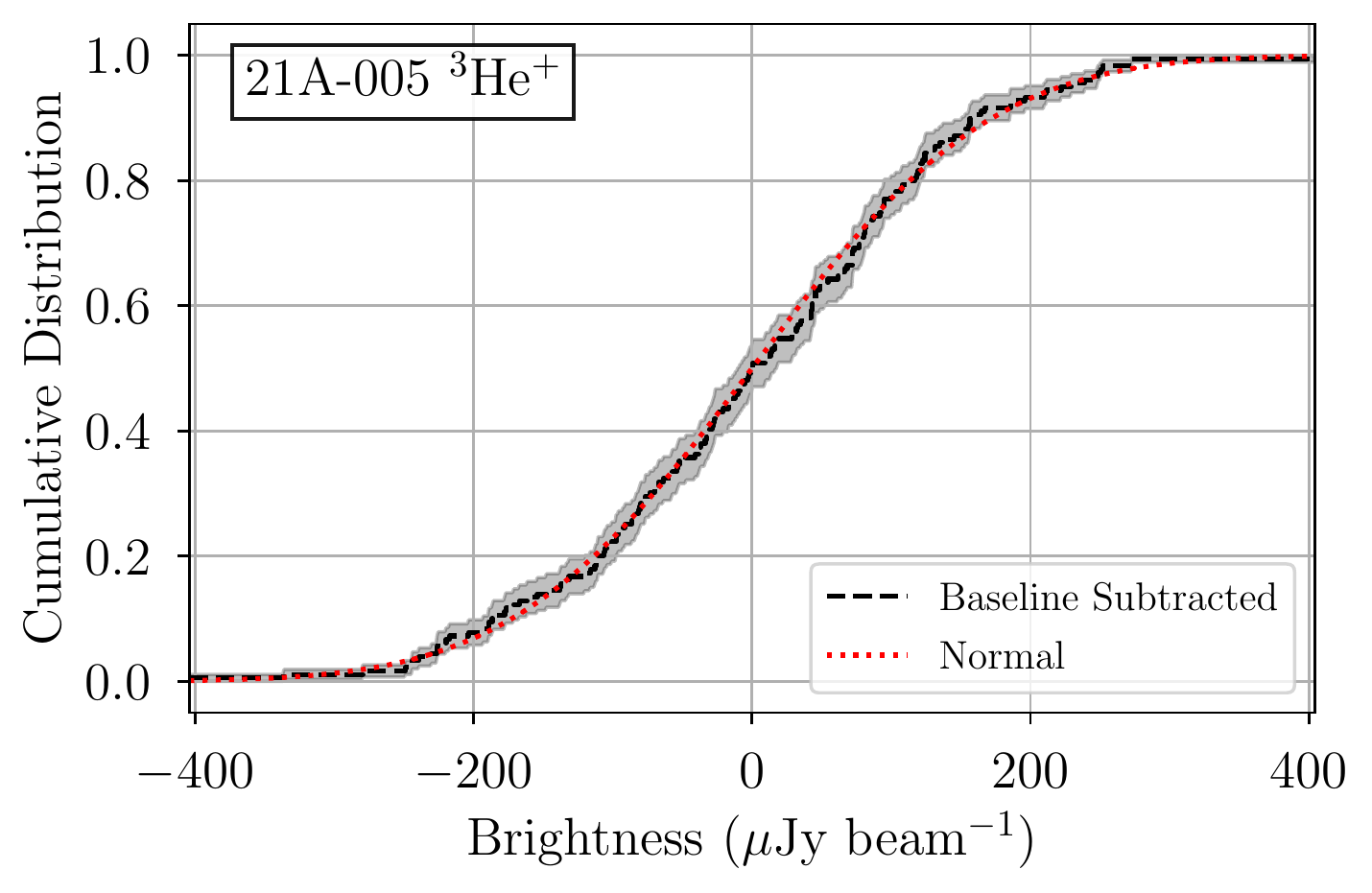}
  \includegraphics[angle=0,scale=0.60]{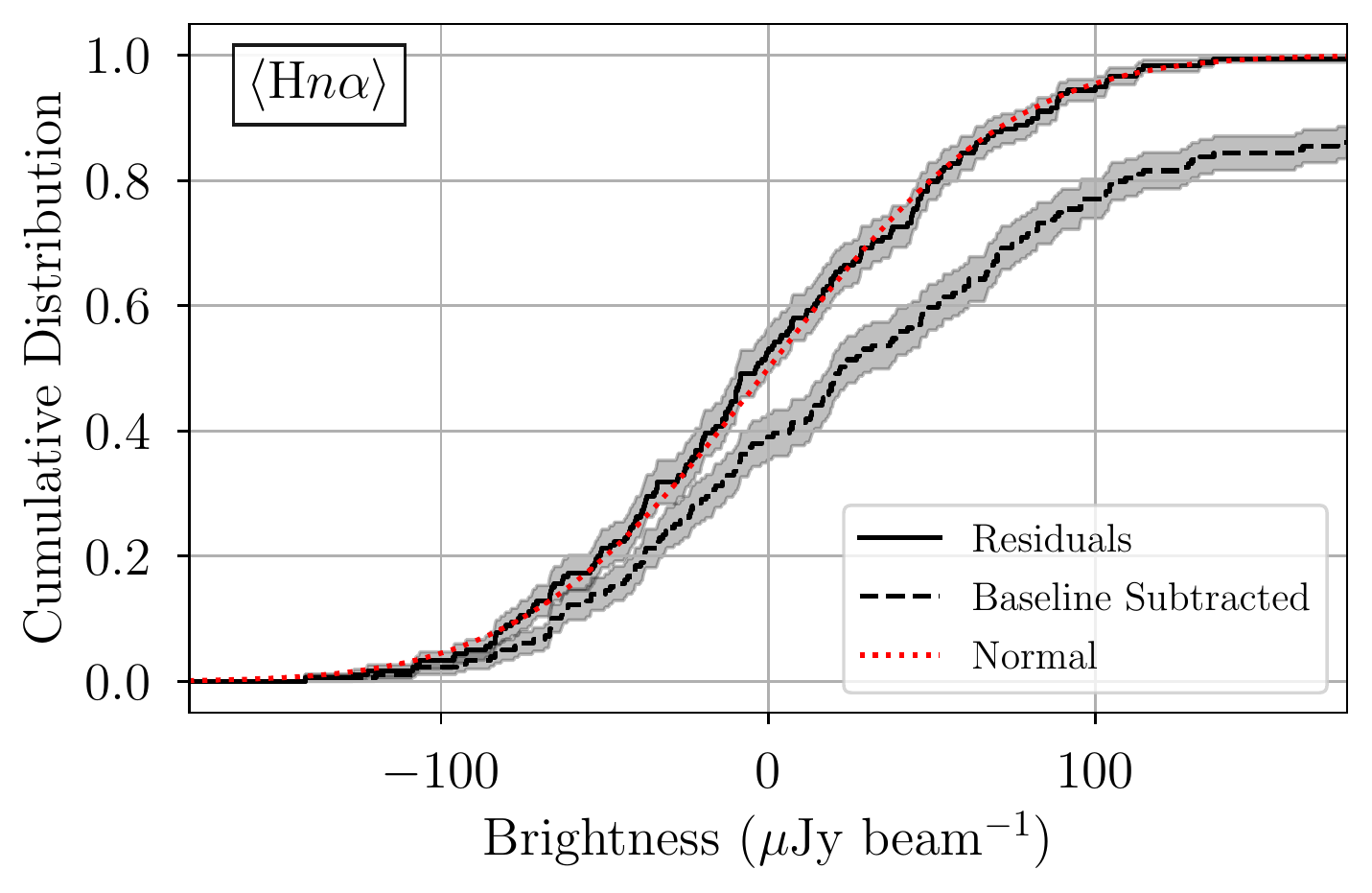}
  \includegraphics[angle=0,scale=0.60]{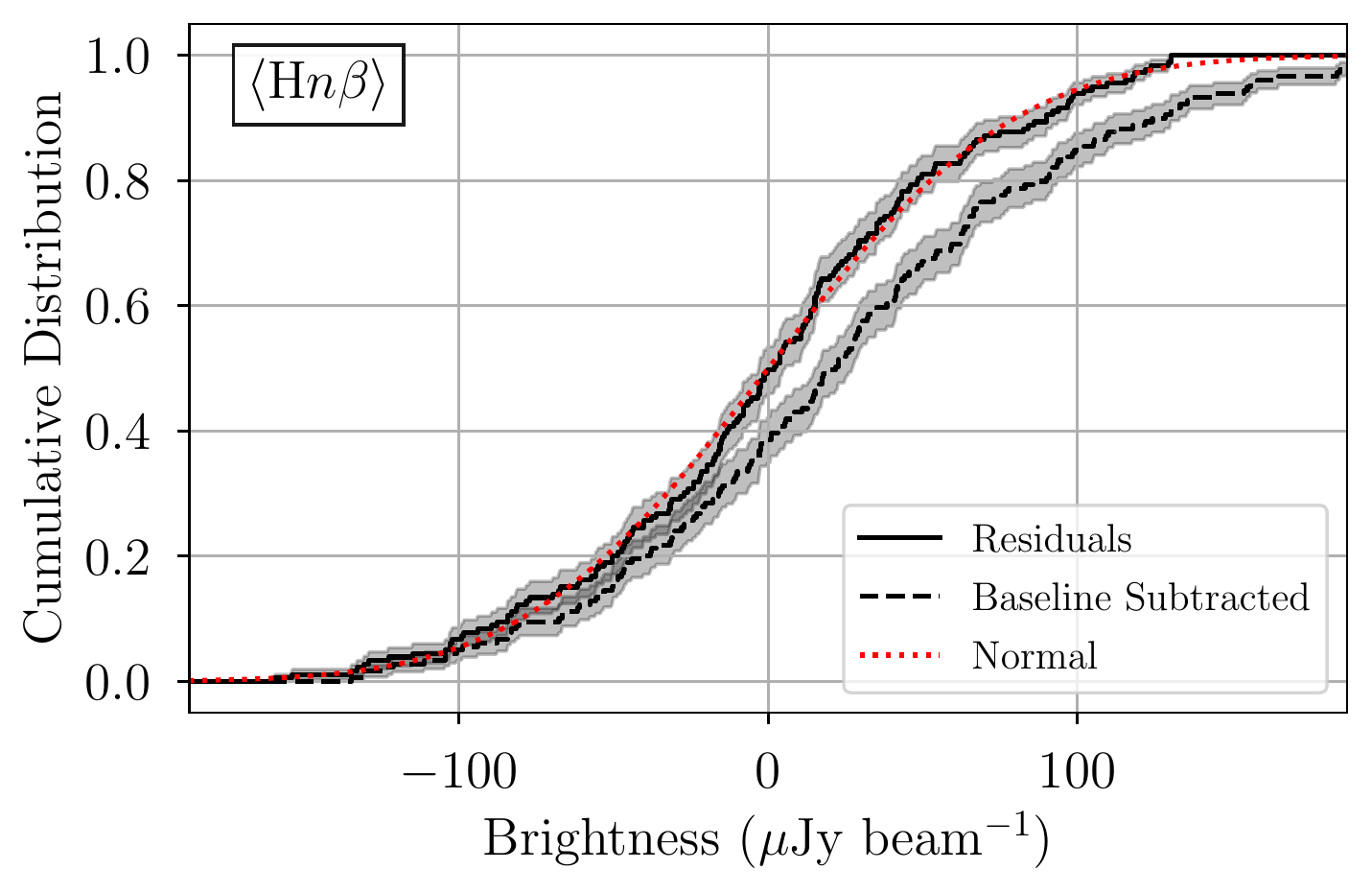}
  \includegraphics[angle=0,scale=0.60]{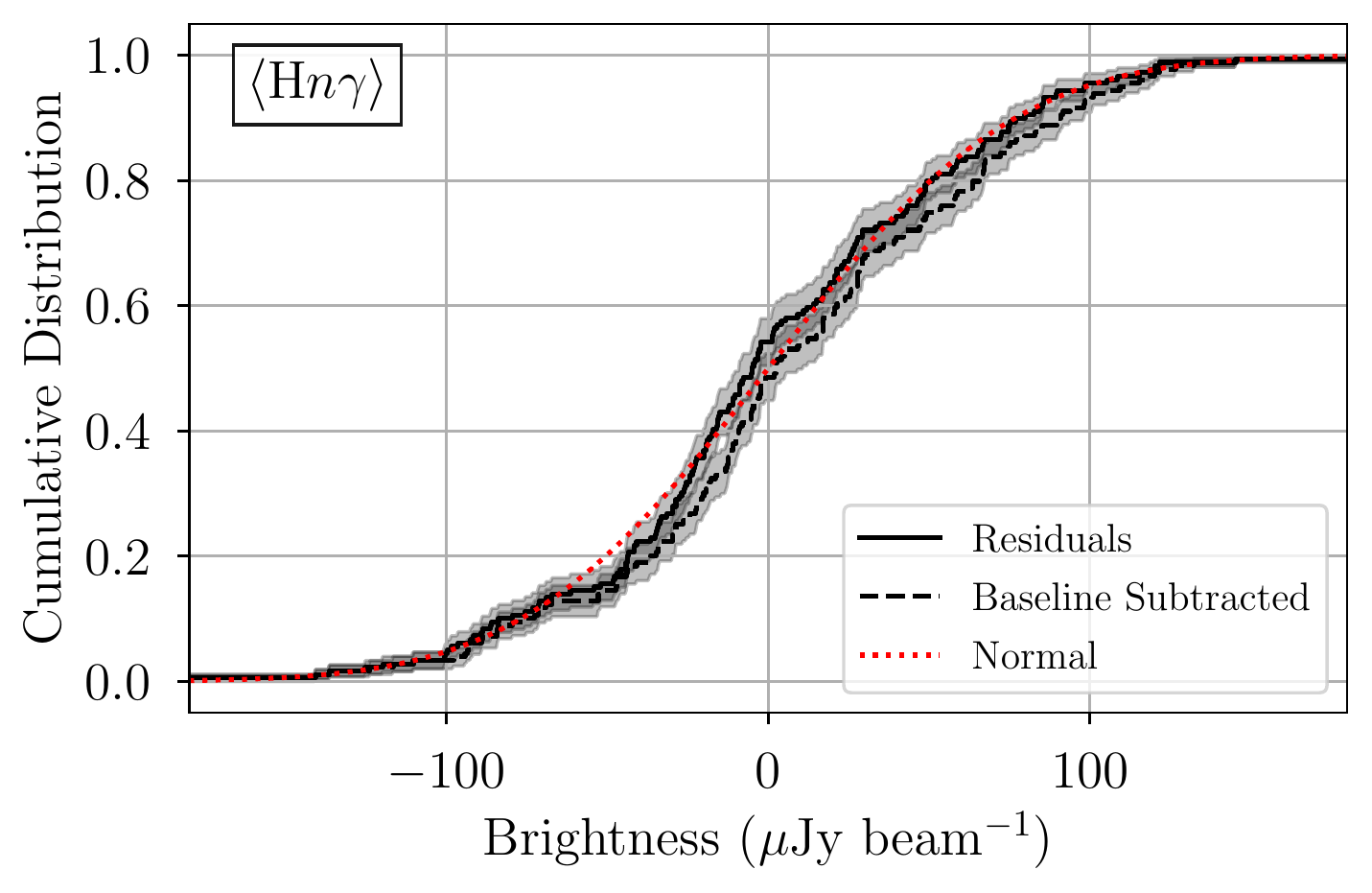}
  \caption{JVLA cumulative distribution functions of the spectra in
    Figure~\ref{fig:spec}: \hep3\ (top-left), \hna\ (top-right), \hnb\
    (bottom-left), and \hng\ (bottom-right).  The dashed black curve
    is the CDF of the baseline subtracted spectrum (middle panel in
    Figure~\ref{fig:spec}); the solid black curve is the CDF of the
    residual spectrum (bottom panel in Figure~\ref{fig:spec}); and the
    red dotted curve is the CDF of theoretical Gaussian noise with an
    RMS given by the residual spectrum.  For the \hep3\ spectrum the
    residual curve is equivalent to the baseline subtracted curve and
    therefore is not shown.  The shaded regions correspond to the 68\%
    confidence interval ($\pm 1\,\sigma$) determined by
    bootstrapping. These figures show significant detections of \hna\
    and \hnb\ emission but only upper limits for \hng\ and \hep3\
    emission.}
\label{fig:cdf}
\end{figure}

First, we compare the cumulative distribution functions (CDFs) of the
\hep3\ and RRL spectra with the CDF expected from theoretical Gaussian
noise.  The results are shown in Figure~\ref{fig:cdf} for the \hep3,
\hna, \hnb, and \hng\ spectra.  Here, the spectra have been smoothed
to 11.4\kms\ and the CDF is calculated over the full spectrum.  To
assess the uncertainty in the CDF distribution we use bootstrap
resampling \citep{efron82}.  Specifically, we generate 10,000
simulated CDFs, where each distribution consists of $M$ values, the
number of data points, that are randomly drawn from the original data
with replacement.  That is, we replace the original data with the
randomly selected data.  So the simulated distributions will miss some
CDF values from the original data and have some duplicates,
triplicates, etc.  The shaded regions in Figure~\ref{fig:cdf}
correspond to the 68\% confidence interval ($\pm 1\,\sigma$)
determined by bootstrapping.  Visual inspection of the CDFs and their
uncertainty indicates that the \hep3\ and \hng\ spectra are consistent
with noise.

To estimate the significance of this result we calculate the {\it
  p}-value, the probability of obtaining a result at least as extreme
as the value of a test statistic.  Here we use the Anderson-Darling
(AD) test statistic which is a weighted sum of the integrated squared
difference between the observed CDF and the theoretical Gaussian CDF
\citep[see][]{scholz87}.  We use the SciPy implementation of the AD
test \citep[see][]{vanderwalt11}.  A significance level threshold of
5\% is typically used \citep[e.g.,][]{feigelson12} and therefore the
spectrum is consistent with random Gaussian noise when the {\it
  p}-value is larger than 0.05.  To do this we run 10,000 simulations
where in each simulation we perform the following steps:
\begin{enumerate}

\item Generate $N$ random observations of a Gaussian distribution, where $N$
  is the same length as the data (or residuals).

\item Calculate the ``nominal'' AD statistic between the data (or
  residuals) and this Gaussian distribution.

\item Generate two bootstrap samples of length $N$ from the combined
  data (or residuals) and random Gaussian observations.

\item Calculate the AD statistic for these two samples and compare
  this to the nominal value in step 2.

\item Calculate the {\it p}-value: the fraction of the time that the
  AD statistic for the two bootstrapped samples is greater than the
  nominal AD statistic.

\end{enumerate}

The results are given in Tables~\ref{tab:he3} and \ref{tab:rrl} for
the \hep3\ spectra and stacked RRL spectra, respectively.  In
Table~\ref{tab:he3} we list the project, the telescope, the velocity
resolution, $\Delta{V_{\rm res}}$, the RMS noise, and the AD {\it
  p}-value.  We also summarize the results for the VLA (AB0794) and the
combined JVLA (21A-004)/VLA (AB0794) data (see below).  The large {\it
  p}-values indicate that the \hep3\ spectrum is consistent with
noise.  This result implies that we have {\it not} detected the \hep3\
transition and that any instrumental spectral baseline effects are
smaller than the expected random Gaussian noise.

In Table~\ref{tab:rrl} we list the RRL order, the velocity resolution,
$\Delta{V_{\rm res}}$, Gaussian fit parameters and S/N for the H and
He RRL components, the RMS noise, and the AD {\it p}-value for the
residual and data spectra.  The Gaussian fit parameters consist of the
peak intensity, $S_{\rm L}$, the full-width half-maximum (FWHM) line
width, $\Delta{V}$, and the LSR velocity of the peak intensity,
$V_{\rm LSR}$.  RRLs are only detected with significance in the \hna\
and \hnb\ spectra.  The residual spectra are consistent with random
Gaussian noise implying that our two-component Gaussian fit is
sufficient.  The signal-to-noise ratios of the He RRLs are less than 2
and therefore we cannot estimate reliable \her4\ abundance ratios.

\begin{deluxetable}{lcccc}
\tabletypesize{\small}
\tablecaption{JVLA (21A-005) and VLA (AB0794) \hep3\ Results \label{tab:he3}}
\tablehead{
  \colhead{} & \colhead{} & \colhead{$\Delta{V_{\rm res}}$} & \colhead{RMS noise\tablenotemark{a}} & \colhead{AD}  \\
  \colhead{Project} & \colhead{Telescope} & \colhead{(\kms)} & \colhead{(\microjyb)} & \colhead{{\it p}-value}
}
\startdata
21A-005 & JVLA      &  2.5 &  134.7 & 0.670 \\
21A-005 & JVLA      & 11.4 &   58.8 & 0.625 \\
AB0794   & VLA       &  8.0 &  127.0 & 0.605 \\
AB0794   & VLA       & 11.4 &  113.3 & 0.600 \\
Combine  & JVLA/VLA  &  8.0 &   68.2 & 0.665 \\
Combine  & JVLA/VLA  & 11.4 &   61.1 & 0.668 \\
\enddata
\tablenotetext{a}{The spectral noise in the data cube.}
\end{deluxetable}

\begin{deluxetable}{lcrccrrccccc} 
\tabletypesize{\tiny} 
\tablecaption{JVLA (21A-005) Radio Recombination Line Results \label{tab:rrl}} 
\tablehead{ 
\colhead{} & \colhead{} & 
\multicolumn{4}{c}{\underline{~~~~~~~~~~~~~~~~~~~~~~~~~~~Hydrogen~~~~~~~~~~~~~~~~~~~~~~~~~~~~}} & 
\multicolumn{3}{c}{\underline{~~~~~~~~~~~~~~~~~~Helium~~~~~~~~~~~~~~~~~~}} & 
\colhead{} & 
\colhead{} & \colhead{} \\ 
\colhead{} & \colhead{$\Delta{V_{\rm res}}$} & 
\colhead{$S_{\rm L}$} & \colhead{$\Delta{V}$} & \colhead{$V_{\rm LSR}$} & \colhead{} & 
\colhead{$S_{\rm L}$} & \colhead{$\Delta{V}$} & \colhead{} & 
\colhead{RMS noise} & 
\multicolumn{2}{c}{\underline{~~~~~AD {\it p}-value~~~~~}} \\ 
\colhead{$\Delta$n} & \colhead{(\kms)} & 
\colhead{(\microjyb)} & \colhead{(\kms)} & \colhead{(\kms)} & \colhead{S/N} & 
\colhead{(\microjyb)} & \colhead{(\kms)} & \colhead{S/N} & 
\colhead{(\microjyb)} & 
\colhead{residuals} & \colhead{data} 
} 
\startdata 
1 & 2.5 & $539.9 \pm  19.8$ & $ 47.9 \pm   2.1$ & $-41.6 \pm   0.8$ & 28.05 & $ 51.9 \pm  22.1$ & $ 38.2 \pm  19.4$ &  2.41 & 59.0 & 0.665 & 0.000 \\ 
1 & 11.4 & $540.3 \pm  23.7$ & $ 48.0 \pm   2.5$ & $-41.6 \pm   1.0$ & 24.98 & $ 52.6 \pm  26.4$ & $ 38.5 \pm  23.1$ &  2.18 & 31.1 & 0.543 & 0.021 \\ 
2 & 2.5 & $163.7 \pm  20.0$ & $ 52.9 \pm   7.8$ & $-40.7 \pm   2.8$ &  8.45 & $ 58.7 \pm  27.6$ & $ 27.0 \pm  15.0$ &  2.16 & 62.4 & 0.622 & 0.010 \\ 
2 & 11.4 & $163.8 \pm  19.4$ & $ 53.1 \pm   7.6$ & $-40.7 \pm   2.7$ &  9.25 & $ 58.0 \pm  26.6$ & $ 27.9 \pm  15.1$ &  2.37 & 26.8 & 0.688 & 0.024 \\ 
3 & 2.5 & $ 48.2 \pm  17.2$ & $ 66.8 \pm  29.2$ & $-40.8 \pm  11.0$ &  2.91 & $  7.2 \pm  28.3$ & $ 23.7 \pm 110.3$ &  0.26 & 60.0 & 0.578 & 0.296 \\ 
3 & 11.4 & $ 48.6 \pm  20.0$ & $ 72.2 \pm  36.3$ & $-35.2 \pm  11.1$ &  2.70 & $ 24.6 \pm  39.5$ & $ 15.0 \pm  28.8$ &  0.62 & 31.7 & 0.597 & 0.398 \\ 
\enddata 
\tablecomments{We follow \cite{lenz92} to calculate the signal-to-noise ratio, S/N.} 
\end{deluxetable}

The spectral noise in the \hep3\ spectral window is consistent with
random Gaussian noise.  Because previous detections of \hep3\ in PNe
have shown to be incorrect \citep{bania21}, we perform several sanity
checks on the data.

\begin{figure}
  \centering
  \includegraphics[angle=0,scale=0.5]{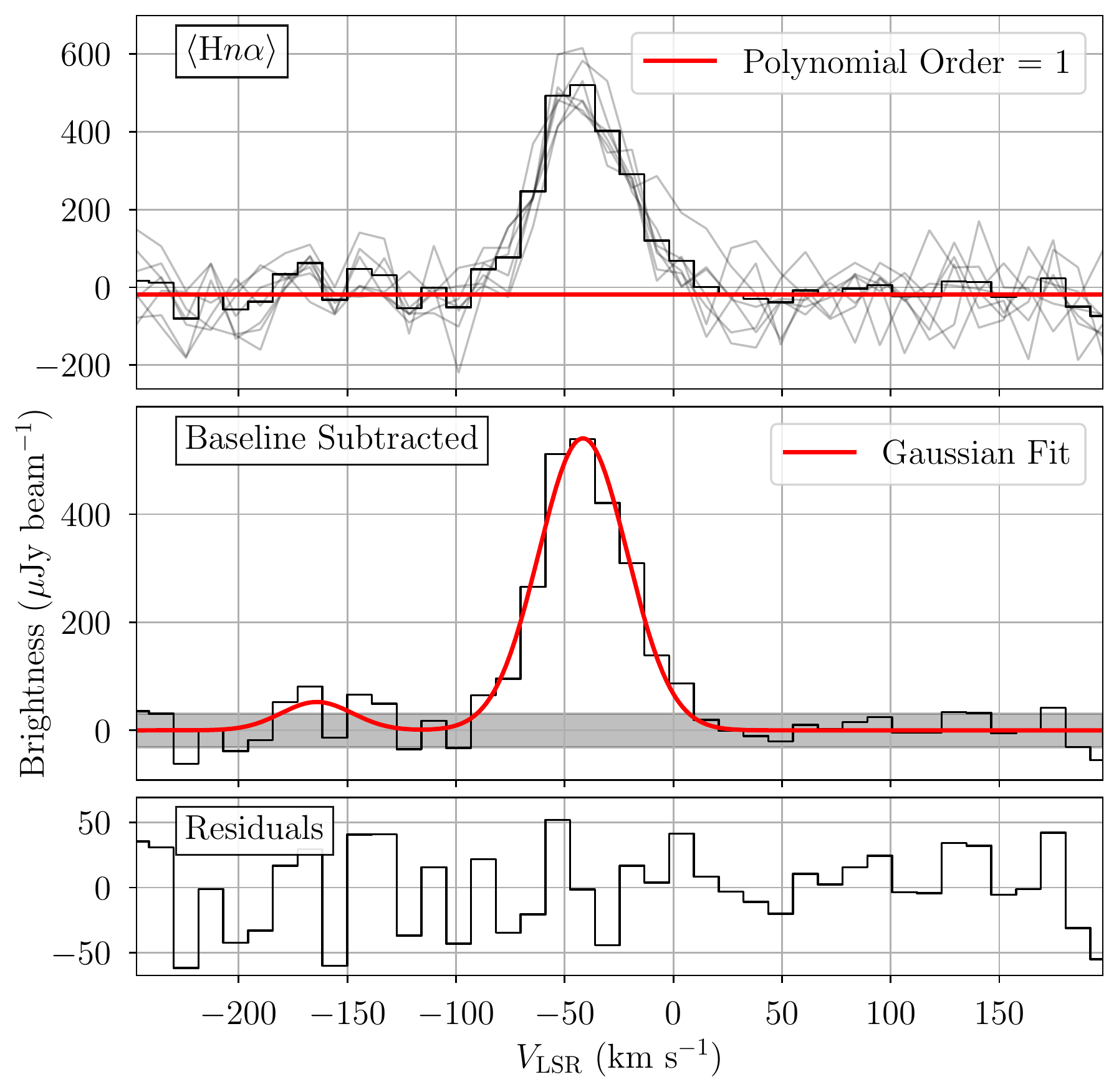}
  \includegraphics[angle=0,scale=0.5]{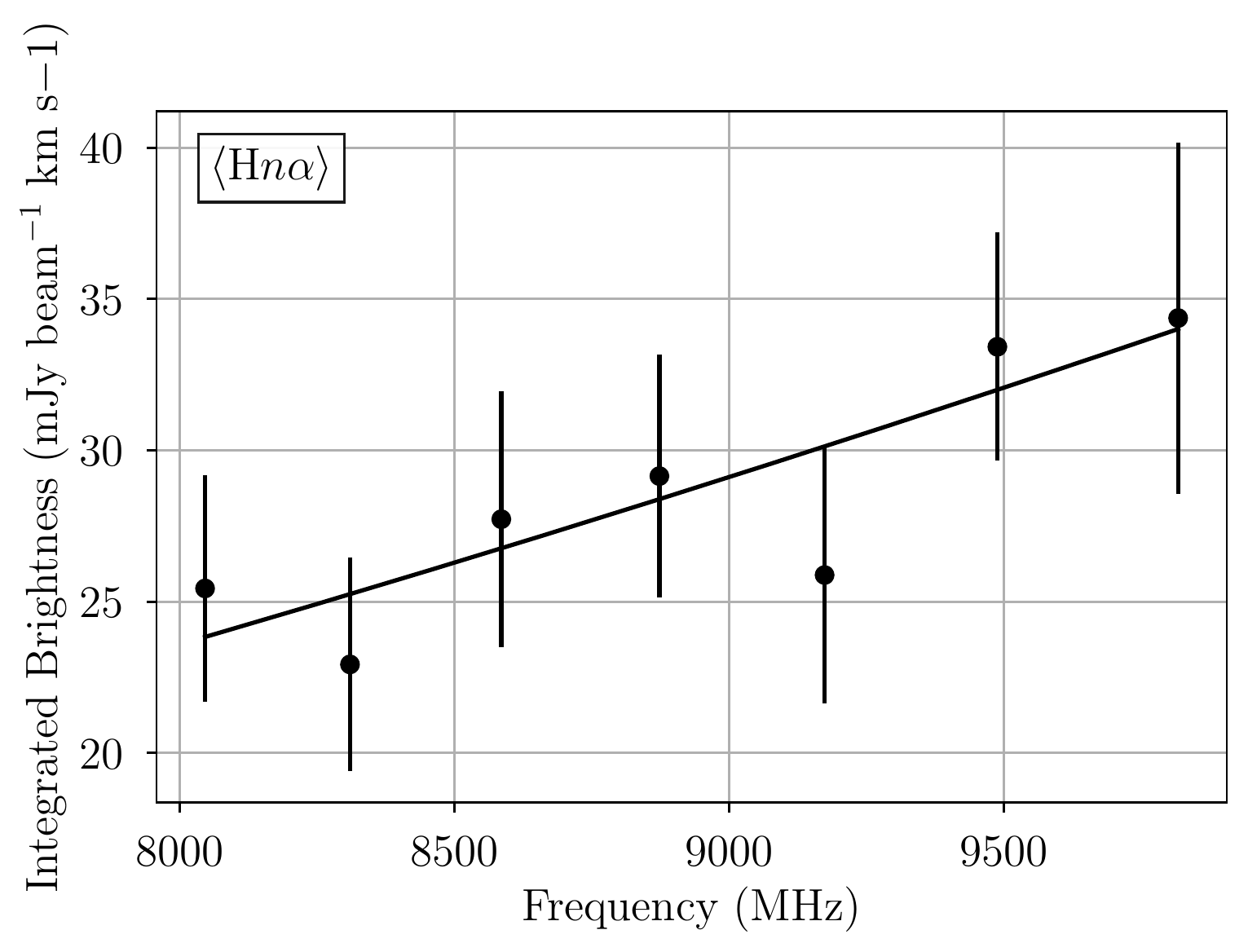} 
  \caption{Analysis of individual Hn$\alpha$ RRL transitions for the
    JVLA data.  {\it Left Panel:} Spectra of 7 RRLs
    (H87$\alpha$--H93$\alpha$) and the stacked \hna\ RRL spectrum in
    bold (top); the \hna\ RRL spectrum after the removal of a spectral
    baseline (middle); and the residuals (bottom).  {\it Right Panel:}
    Integrated Hn$\alpha$ brightness as a function of frequency.  The
    solid line is a power-law fit to the points with an exponent of
    $1.79 \pm 0.84$.}
\label{fig:sed}
\end{figure}

\begin{enumerate}

\item {\it Does the noise integrate down as expected?}.  The JVLA
  exposure calculator tool\footnote{See go.nrao.edu/ect.} predicts an
  RMS noise of 51.5\microjyb, assuming a total time of 29\hr\ and a
  velocity resolution of 11.4\kms.  Based on the number of baselines
  flagged we estimate an effective integration time of 18.8\hr,
  increasing the RMS noise to 64.0\microjyb.  This is roughly
  consistent with our measured value of 58.8\microjyb. 

\item {\it Do adjacent RRLs behave as expected?} Because RRLs at
  centimeter wavelengths have large principal quantum numbers, the
  energy spacing between adjacent RRLs is similar and therefore these
  transitions should have similar RRL spectral properties.  Moreover,
  for an optically thin nebula we expect the integrated RRL flux
  density to increase with frequency: $\int S_{\nu}\,d\nu \propto \nu$
  \citep{wilson12}. Figure~\ref{fig:sed} shows that both of these
  expectations are true for the Hn$\alpha$ RRLs in J320.  Individual
  Hn$\alpha$ RRLs have similar profiles and the integrated flux
  density increases linearly with frequency to within the uncertainty.
  A power-law fit to the data yields an exponent of $1.79 \pm 0.84$.
  This is a large deviation from the expected exponent of 1.0 for an
  optically thin nebula, but consistent to within the uncertainties.
  
\item {\it Are the RRLs in LTE?} There have been non-LTE effects
  detected for RRLs in PNe, but in general we expect centimeter
  wavelength RRLs to be close to LTE \citep[e.g., see][]{bania21}.  In
  LTE we expect the following integrated intensity ratios:
  H114$\beta$/H91$\alpha$ = 0.274 and H130$\gamma$/H91$\alpha$ = 0.126
  \citep{bania21}.  From Table~\ref{tab:rrl} we measure
  $\langle {\rm Hn}\beta \rangle/\langle {\rm Hn}\alpha \rangle =
  0.335 \pm 0.067$ and
  $\langle {\rm Hn}\gamma \rangle/\langle {\rm Hn}\alpha \rangle =
  0.125 \pm 0.071$.  So within the uncertainties the RRL emission is
  consistent with LTE excitation in J320.

\end{enumerate}

Many PNe, including J320, contain diffuse halos that can be detected
with deep H$\alpha$ observations.  Since the intensity of free-free
and RRL emission is proportional to $\int n_{\rm e}^{2}\,d\ell$, these
tracers are not a very sensitive probe of the halo.  In contrast, the
intensity of the \hep3\ transition is proportional to
$\int n_{\rm e}\,d\ell$ and thus the halo could contribute
significantly to the \hep3\ emission \citep[see][]{balser97}.  To
increase our sensitivity to \hep3\ emission, we therefore follow
\citet{balser06b} and integrate \hep3\ emission over both the line
profile and spatially around J320.  Specifically, we first integrate
over the expected FWHM line width of the \hep3\ transition; that is,
we produce a channel-integrated image from the \hep3\ data cube.
Using this image we then spatially integrate over concentric rings
centered on J320.

\begin{figure}
  \includegraphics[angle=0,scale=0.55]{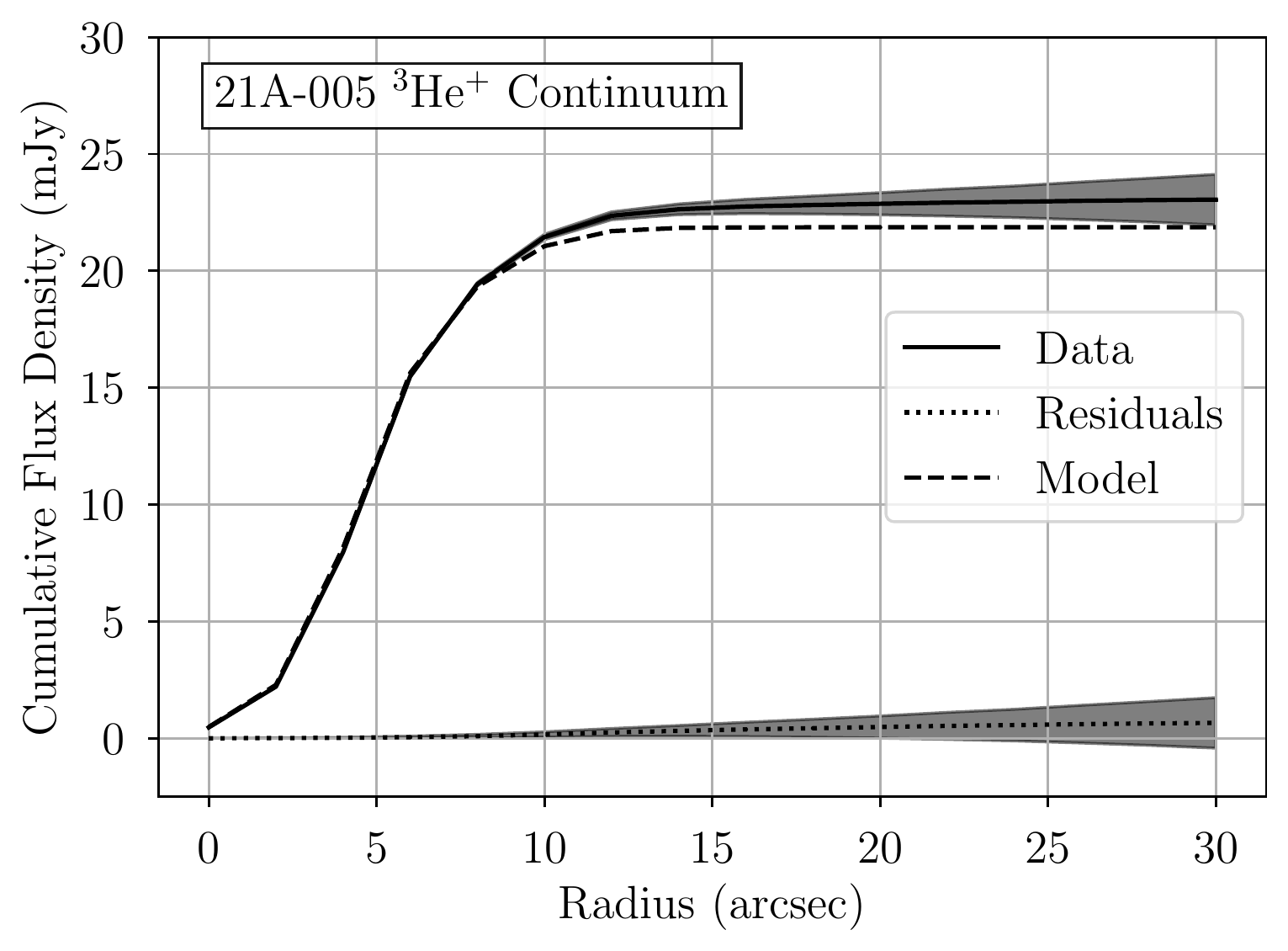} 
  \includegraphics[angle=0,scale=0.55]{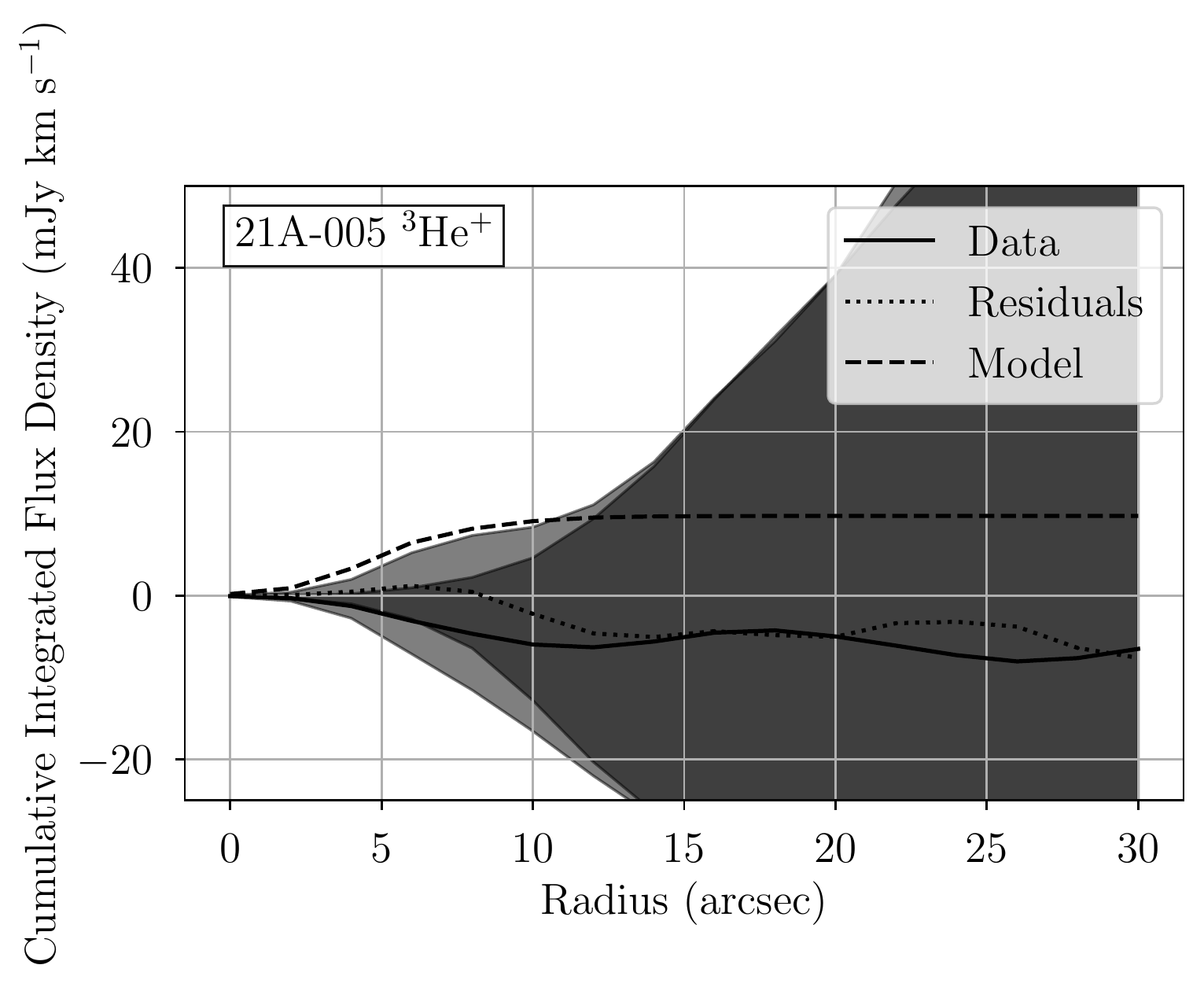}
  \caption{Spatially integrated free-free continuum (left) and \hep3\
    line (right) flux densities as a function of radius for the JVLA
    (21A-005) data.  For the \hep3\ line emission we first integrate
    over spectral channels and then spatially around J320.  {\it Data
      (solid curves):} The continuum MS-MFS image of the \hep3\
    spectral line window (left) and the channel-integrated \hep3\
    spectrum (right).  Both images were smoothed to 12\arcsec\
    resolution.  {\it Residuals (dotted curves):} The continuum
    residual cumulative flux density is measured in the MS-MFS
    residual image (left), and the \hep3\ spectral line residual
    cumulative integrated flux density is measured in the line-free
    channels (right).  The shaded regions represent the $1\,\sigma$
    uncertainties determined from the residual (continuum) or
    line-free channels (\hep3\ spectral line).  {\it Model (dashed
      curves):} Results of a similar analysis for the NEBULA model
    free-free continuum image and \hep3\ spectral line cube (see
    Section~\ref{sec:model}).}
\label{fig:iring}
\end{figure}

Figure~\ref{fig:iring} shows the results of this analysis for the JVLA
data.  In the left panel we plot the cumulative free-free continuum
flux density as a function of radius.  The free-free emission levels
off beyond about 15\arcsec; that is, we do not detect any halo
emission with our radio data.  This is expected given the low emission
measure probed by H$\alpha$ emission in PNe halos.  In the right panel
we plot the cumulative, channel-integrated \hep3\ line flux density as
a function of radius.  No \hep3\ emission is detected.  The noise,
shown by the shaded regions for the data and residuals, increases with
radius since we are integrating over a noisy signal.  The darker
shaded region is where the noise envelopes of the data and residual
curves overlap.

\begin{figure}
  \includegraphics[angle=0,scale=0.45]{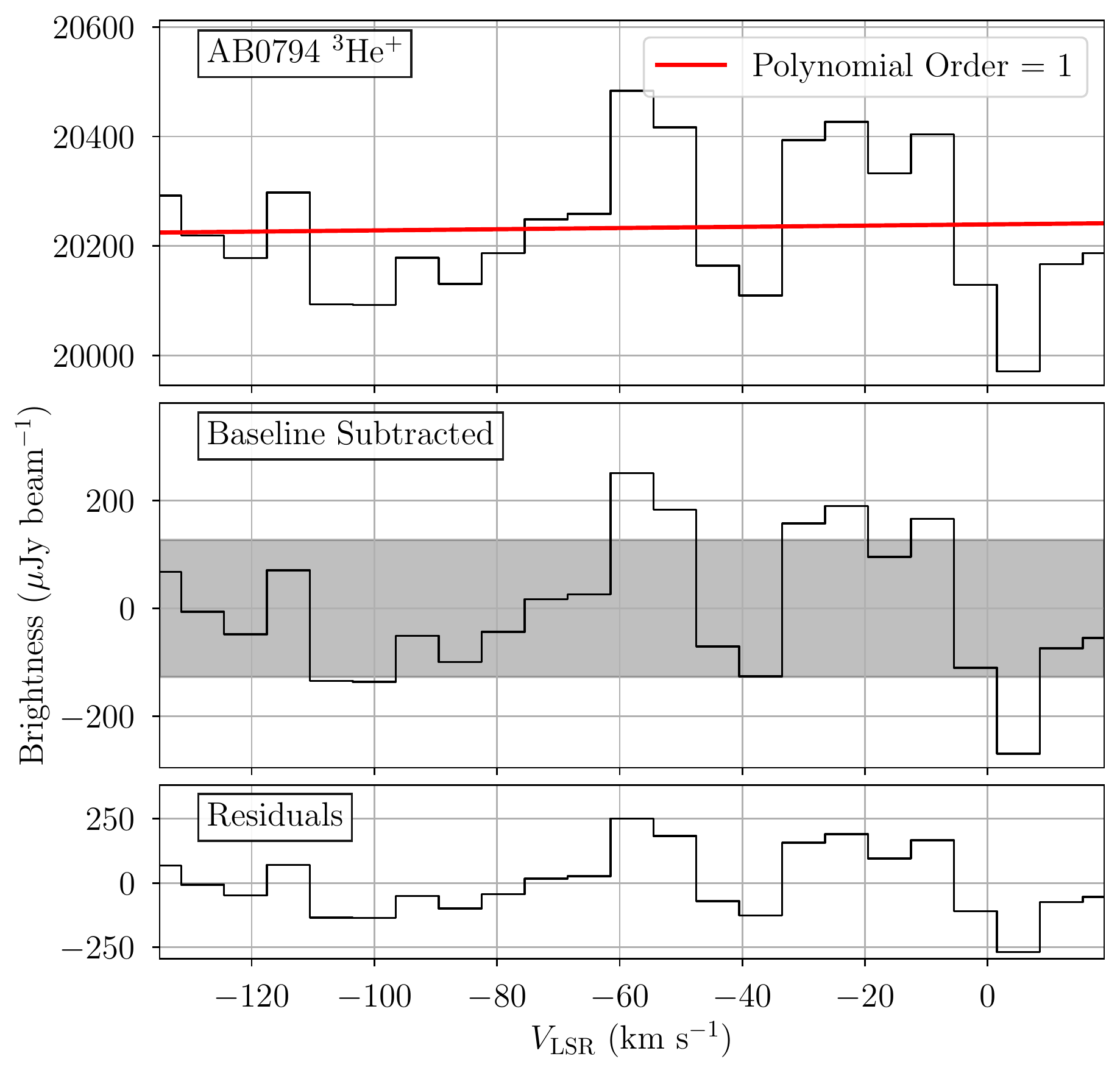}
  \includegraphics[angle=0,scale=0.45]{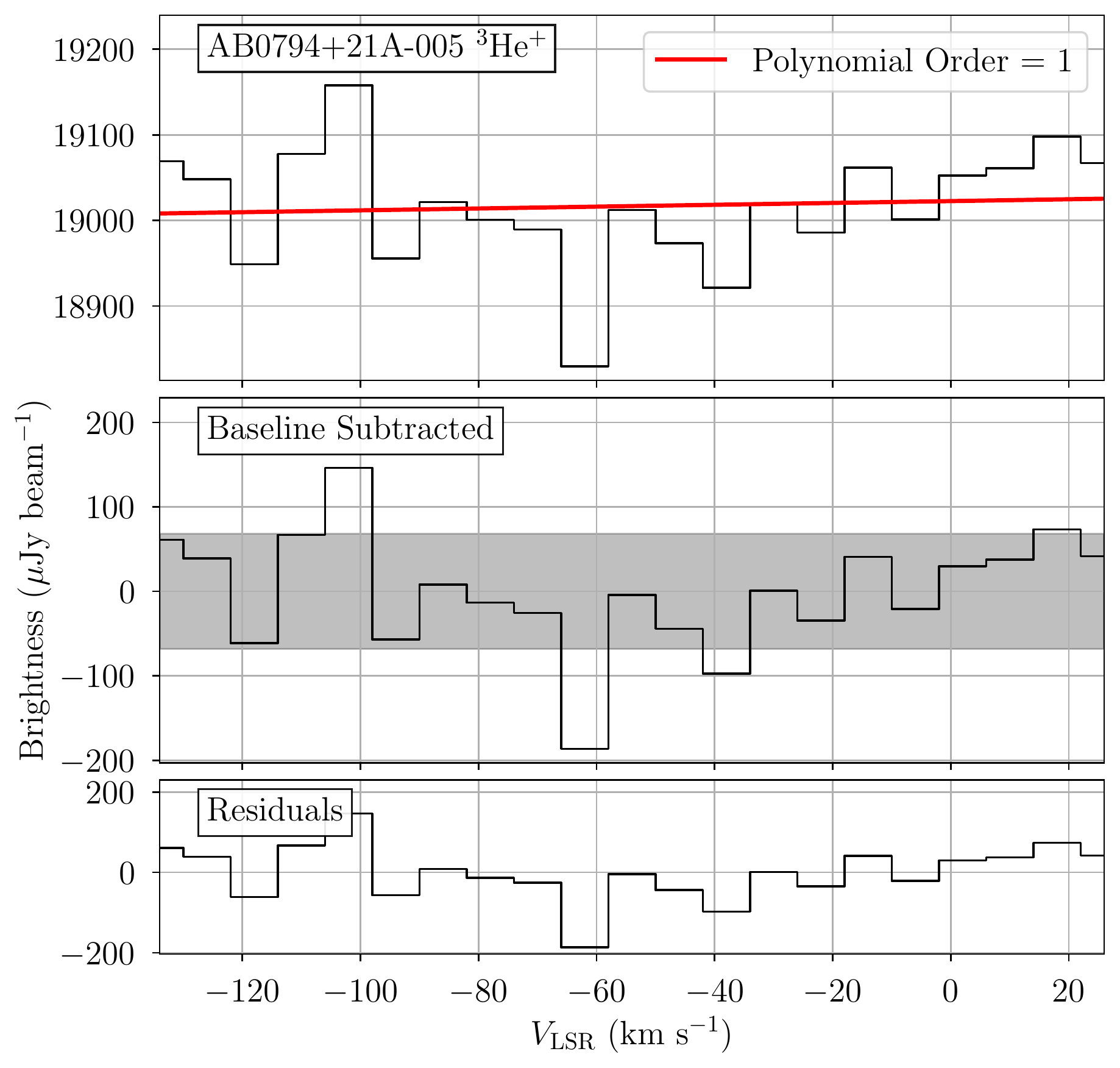}
  \includegraphics[angle=0,scale=0.6]{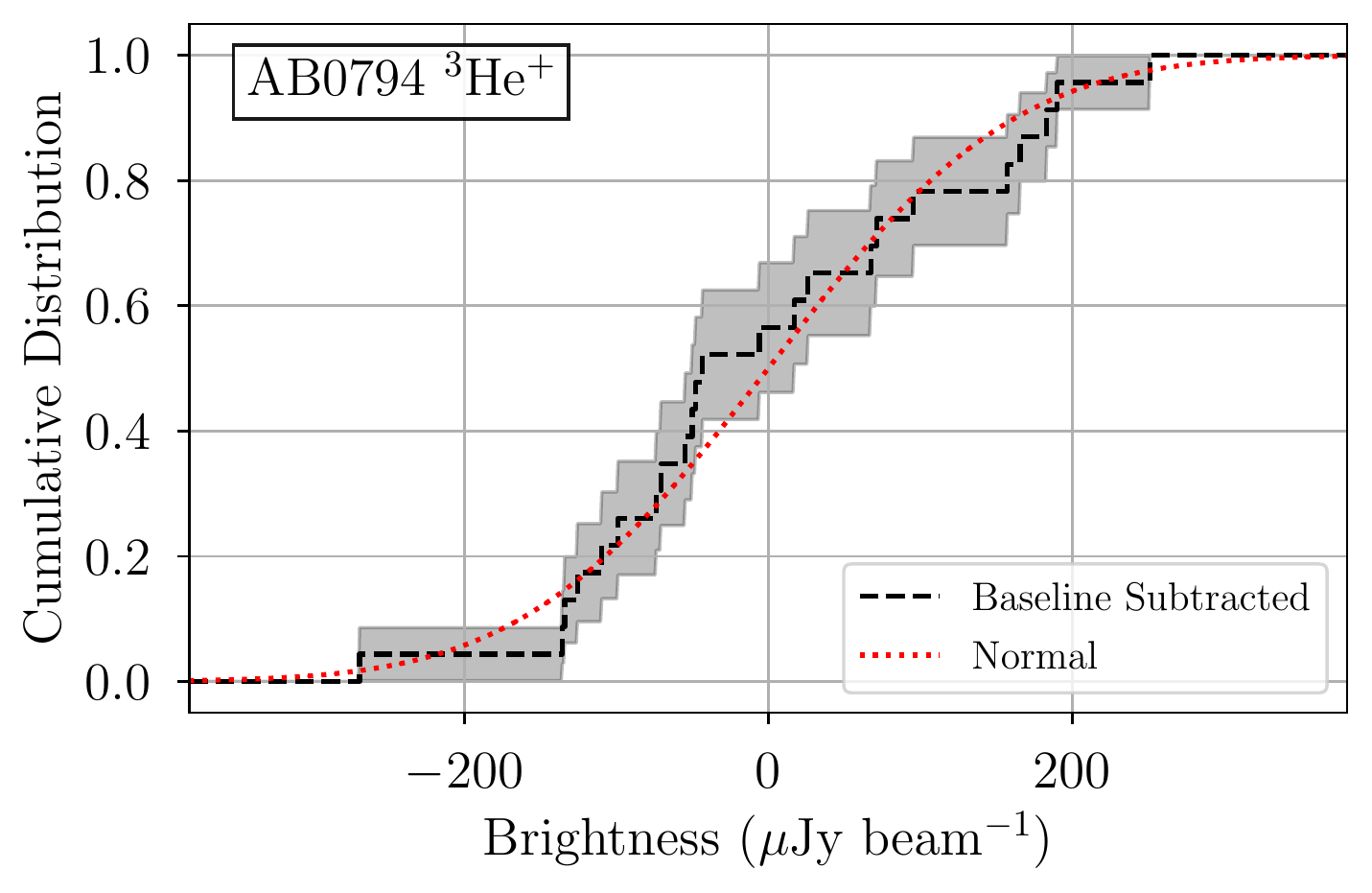} \hspace{0.6cm}
  \includegraphics[angle=0,scale=0.6]{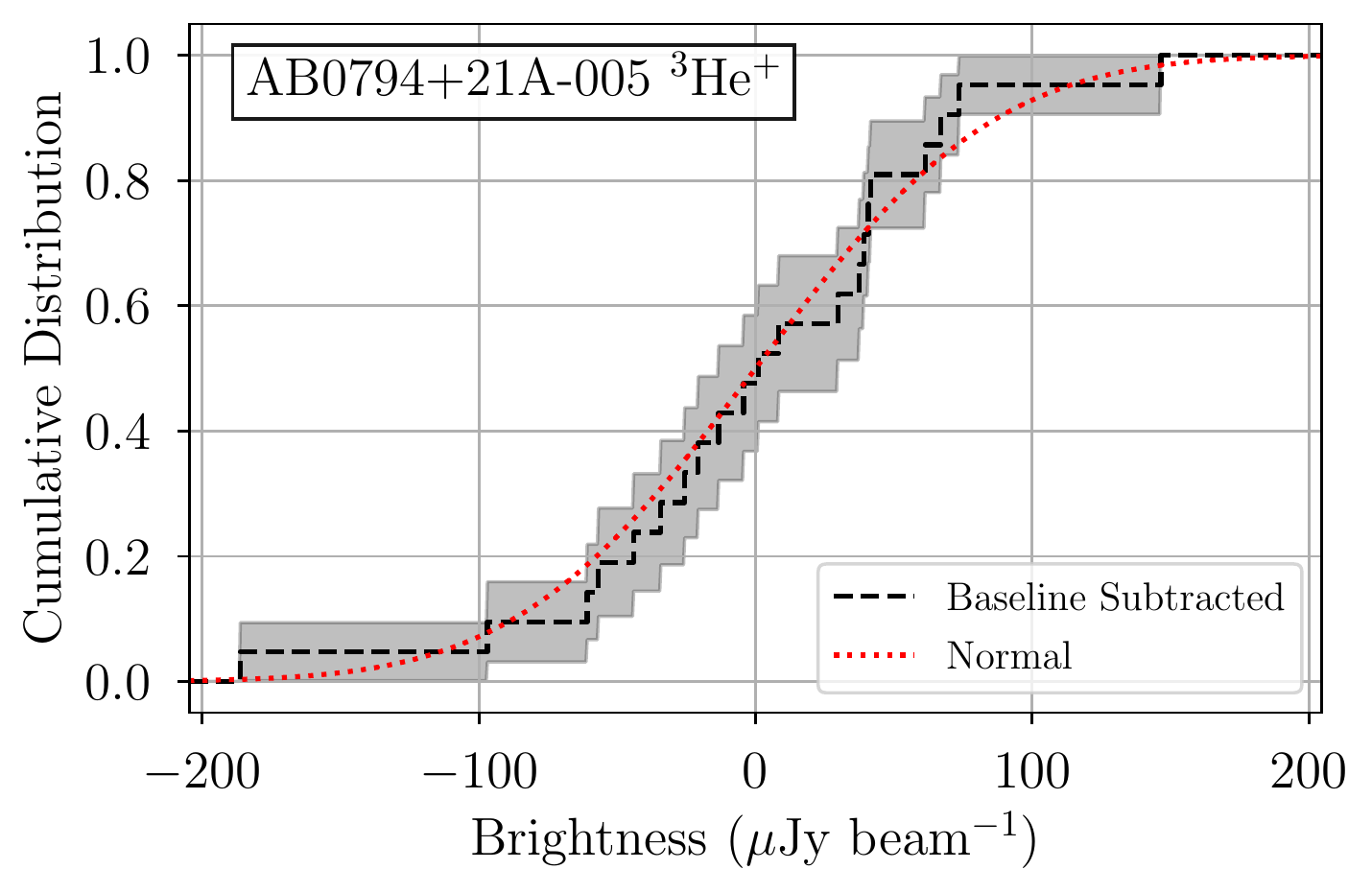}
  \caption{J320 \hep3\ spectra (top) and CDFs (bottom) using the VLA
    (project AB0794; left) and the combined VLA and JVLA data
    (projects AB0794 and 21A-005; right).  Spectra are smoothed and
    regridded to a velocity resolution of 8.0\kms.  See
    Figures~\ref{fig:spec} and \ref{fig:cdf} for details.  Combining
    the VLA and JVLA J320 data does not lead to a \hep3\ detection.}
\label{fig:old}
\end{figure}

To compare our results with \citet{balser06b}, we reanalyze their VLA
data (project AB0794) using WISP with the same procedures as for our
JVLA data for consistency.  We also combine the VLA and JVLA data to
increase our sensitivity.  Since the VLA correlator was limited to
spectral windows with 31 channels over a 6.25\mhz\ bandwidth, we first
smooth our JVLA data to the same spectral resolution of 8.0\kms.
Spectra and CDFs are shown in Figure~\ref{fig:old}.  There is a hint
of a double-peaked \hep3\ profile in the reprocessed VLA data which is
consistent with the results in \citet[][Figure~7]{balser06b}.  But
this feature is not present in the JVLA data nor in the combined data
set.  Moreover, all spectra are consistent with random Gaussian noise
(see the CDFs in Figure~\ref{fig:old} and the AD {\it p}-values in
Table~\ref{tab:he3}).

The spectral line RMS noise in the reprocessed VLA data is almost two
times higher than the JVLA data using the coarse velocity resolution
data cubes with $\Delta{V_{\rm res}} = 11.4$\kms\ (see
Table~\ref{tab:he3}).  There are three factors that account for most
of this difference.  (1) The lower system noise in the JVLA receivers
produces a system equivalent flux density (SEFD) of $\sim 250$\jy\ at
X-band compared to $\sim 310$\jy\ for the VLA.  (2) The JVLA WIDAR
correlator efficiency is 0.93 for the 8-bit samplers, whereas the
3-level VLA correlator had an efficiency of 0.78. (3) The VLA data
require significant flagging due to bad data resulting in an effective
integration time of 13.5\hr, whereas the cleaner JVLA data set has an
effective integration time of 18.8\hr.  In total these three
differences account for a factor of 1.7 in RMS sensitivity between the
VLA and JVLA data sets.  Thus combining the VLA data with the JVLA
data does not significantly change the measured RMS noise.

{\it Why did \citet{balser06b} claim a $^{\it 3}$He$^{\it +}$
  detection?}  Their \hep3\ spectrum toward the peak continuum
emission is consistent with our results: no significant \hep3\
detection.  The difference arises when generating the cumulative,
channel-integrated flux density where they achieve a S/N of $\sim 9$
(cf., Figure~8 in \citet{balser06b}).  The main reason for the
discrepancy stems from performing a continuum subtraction using the
few available channels in the VLA correlator to define the line-free
regions.  This produces a poor spectral baseline fit that artificially
amplifies the ``noise'' bumps visible in the single spectrum centered
on J320.  The magnitude of the \hep3\ signal increases when
calculating the cumulative, channel-integrated flux density.  In
contrast, the JVLA data have ample channels to define the line-free
region, and moreover since the line data are not limited by dynamic
range a continuum subtraction is not necessary.  By processing the VLA
data in the same way as \citet{balser06b} we can reproduce their
results.

\subsection{J320 NEBULA Model}\label{sec:model}

Deriving a limit for the \her3\ abundance ratio in J320 requires a
model of the nebular structure for this PN.  \citet{balser06b} modeled
J320 as a two-component nebula consisting of a bright shell and a
diffuse, extended halo.  They used the numerical radiative transfer
code NEBULA \citep{balser95, balser99a, balser18} to calculate
synthetic spectra of the free-free continuum, RRL, and \hep3\ emission
from this model nebula.  Specifically, NEBULA produces a continuum
image and spectral line data cubes.

Here, we use more recent optical and infrared data together with our
sensitive JVLA radio data to constrain the physical properties of
J320.  We adopt a distance of 3.26\kpc\ based on parallax observations
from Gaia DR2 \citep{chornay20}.  J320 is morphologically complex but
there exists a brighter shell with an angular diameter of
$\sim 7$\arcsec\ embedded within a larger halo of $\sim 25$\arcsec\
\citep{harman04}.  The angular size of the inner boundary of the shell
is difficult to discern from the optical images so we assume
$0\arcsper1$.  The helium ionization structure is taken from optical
recombination lines where we assume the doubly ionized helium is
negligible \citep{costa04}.  For an optically thin nebula in LTE the
radio recombination line-to-continuum ratio is an accurate measure of
the electron temperature that is independent of density \citep[e.g.,
see][]{wenger19b}.  Using the H91$\alpha$ RRL we derive
$T_{\rm e} = 12,500$\K, consistent with results derived from optical
collisionally excited lines \citep[e.g.,][]{milingo02, costa04}.
Using infrared collisionally excited lines of sulfur
\citet{pagomenos18} derive an electron density of
$n_{\rm e} = 3,350 \pm 600$\percc.  This produces radio continuum
emission brighter than observed with the JVLA so we reduce this value
to $n_{\rm e} = 2,100$\percc\ to be consistent with our radio
observations.  The infrared and radio data are not probing the same
volume for this complex nebula, therefore reducing the density is
justified.

The RRL line widths in J320 are broadened by Doppler motions,
consisting of both thermal and non-thermal components, together with
expansion of the nebula.  \citet{harman04} measured bipolar lobes
expanding with a velocity of $V_{\rm exp} = 46$\kms, but this value is
unlikely to be representative of our RRL emission which arises from
the entire volume of ionized gas.  We therefore adopt an expansion
velocity of $V_{\rm exp} = 16$\kms\ based on H$\alpha$ emission of the
entire shell \citep{rechy-garcia20}.  The thermal motions are
determined by the electron temperature which we derive to be
$T_{\rm e} = 12,500$\K\ (see above).  The non-thermal motions, thought
to be caused by turbulence, are constrained by the observed JVLA line
widths.  That is, we increase the model turbulent velocity until the
synthetic RRL line widths are consistent with the observed line
widths.

The NEBULA model physical parameters for J320 are summarized in
Table~\ref{tab:model}.  Listed are the nebular component, the inner
and outer angular sizes, the expansion velocity, the electron
temperature, the electron density, and the helium ionic abundances.
The physical parameters of the halo are not well determined because
the low density produces weak emission lines.  We therefore assume
that the halo has the same expansion velocity, electron temperature,
and helium ionization structure as the shell.  \citet{balser06b} used
the \hep3\ and continuum emission distribution to constrain the halo
density and \her3\ abundance ratio.  Since we do not detect \hep3\
with the improved JVLA observations, we arbitrarily set the density to
a low value of $n_{\rm e} = 10$\percc.  A \her3\ abundance ratio of
$\nexpo{2.75}{-3}$ by number produces a limit to our JVLA observations
(see below).

\begin{figure}
  \centering
  \includegraphics[angle=0,scale=0.55]{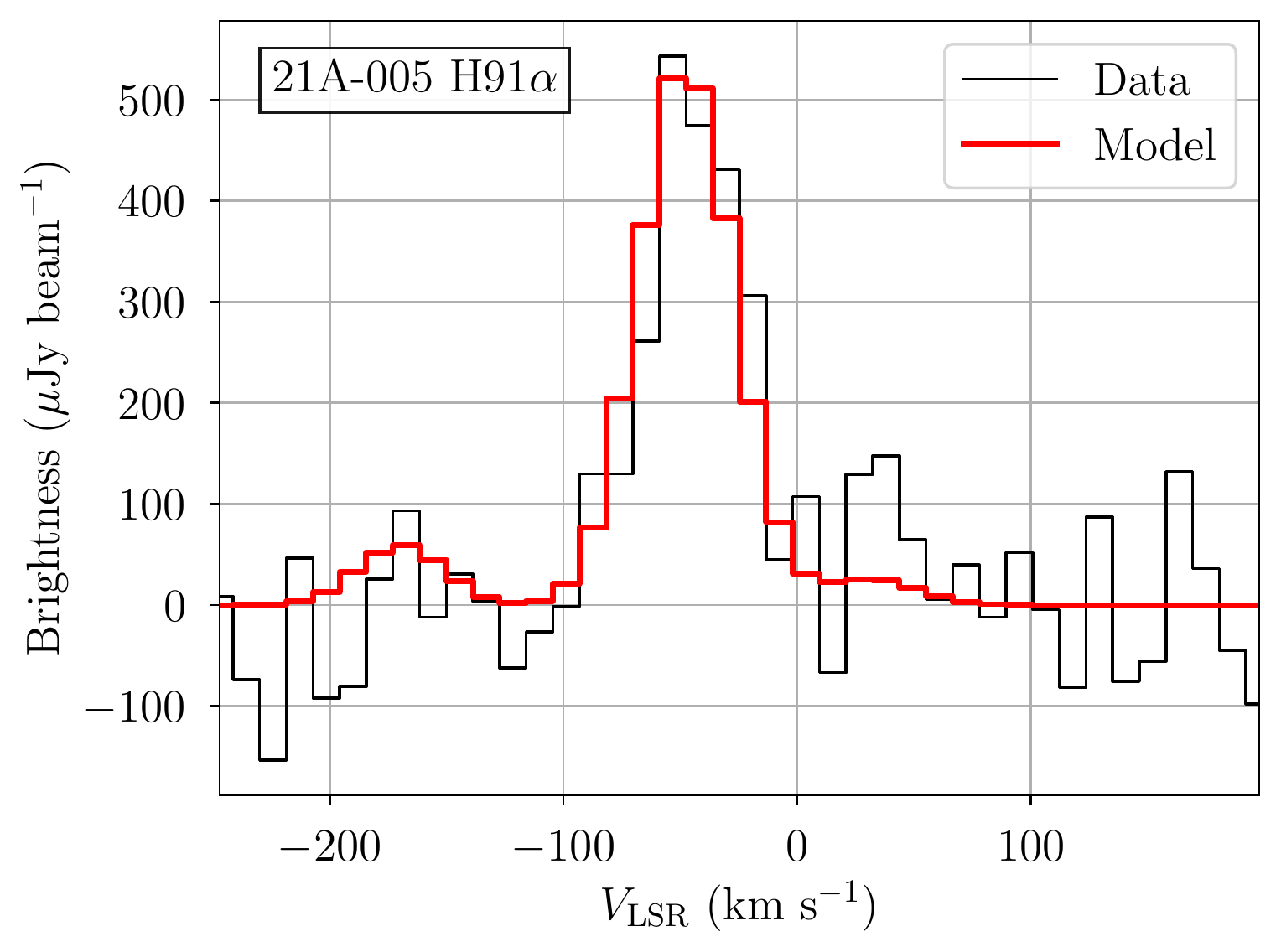}
  \includegraphics[angle=0,scale=0.55]{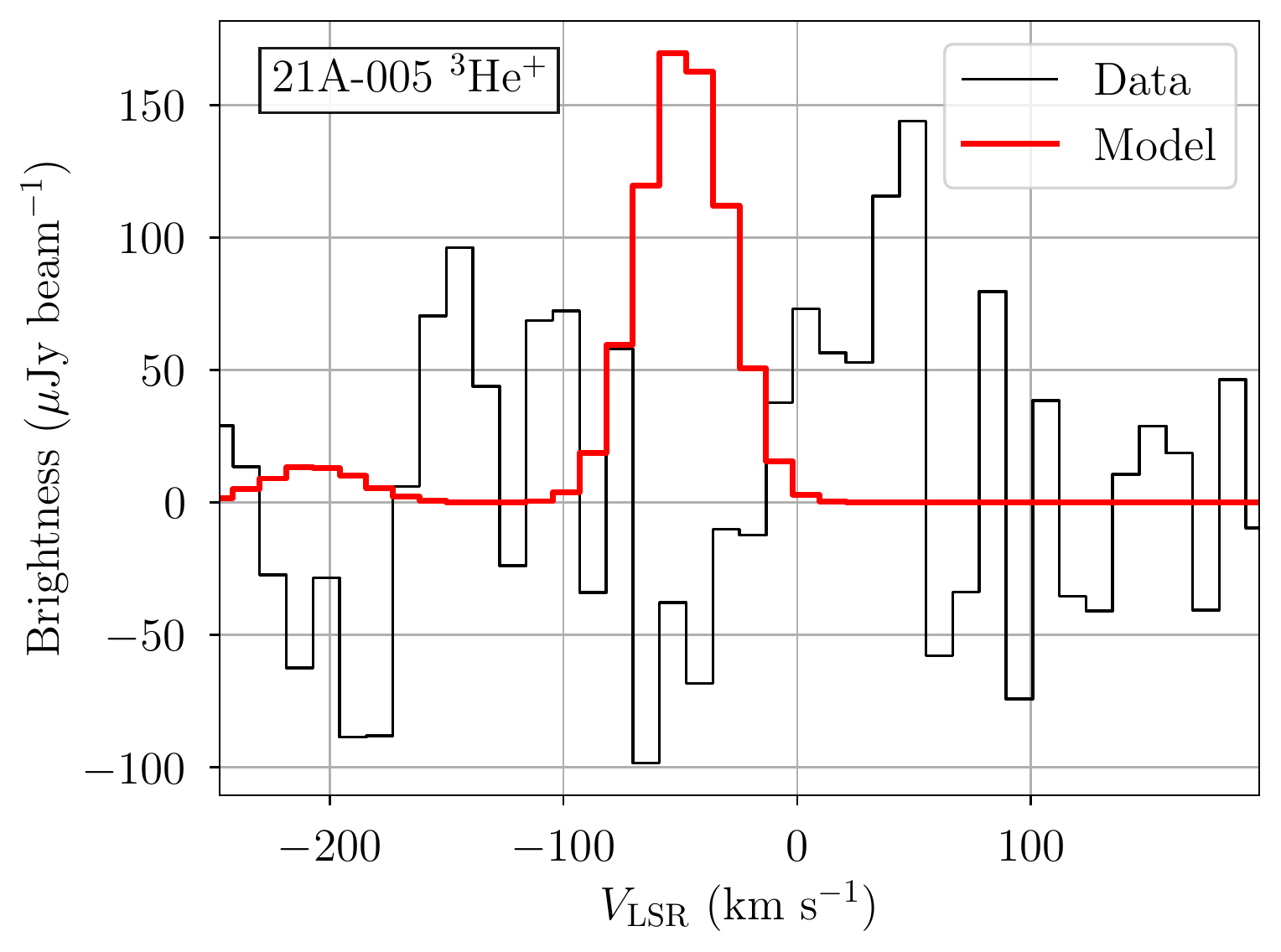}
  \caption{J320 NEBULA model synthetic spectra of the H91$\alpha$ RRL
    and the \hep3\ transition.  JVLA (project 21A-005) spectra are
    shown as black histograms for comparison.  The data and model
    spectral line cubes have been smoothed to a 12\arcsec\ angular
    resolution.  {\it Left Panel:} H91$\alpha$ RRL band.  The model
    H91$\alpha$ RRL is a reasonably good fit to the data.  The
    He91$\alpha$ RRL (near $V_{\rm LSR} \sim -165$\kms) and
    H154$\epsilon$ RRL (near $V_{\rm LSR} \sim 26$\kms) are included
    in the NEBULA model but are too weak to be detected with the JVLA.
    {\it Right Panel:} \hep3\ transition band.  The model \hep3\
    intensity is about 2--3 times the JVLA RMS noise (see
    Table~\ref{tab:he3}).  The H171$\eta$ RRL (near
    $V_{\rm LSR} \sim -210$\kms) is included in the model but too weak
    to be detected with the JVLA.}
\label{fig:model}
\end{figure}

We analyze the NEBULA data products of J320 using the same methods as
for our JVLA data.  The model brightness distribution is convolved
with a Gaussian beam with HPBW of 12\arcsec.  Spectra of the
H91$\alpha$ RRL and \hep3\ transition toward the peak continuum
emission in J320 are shown in Figure~\ref{fig:model}.  The synthetic
H91$\alpha$ profile is a reasonable fit by eye to the JVLA data.  This
is expected since the model electron temperature and density were
constrained using the JVLA H91$\alpha$ line-to-continuum ratio and the
free-free radio continuum emission centered on J320.  The NEBULA model
produces a \hep3\ line intensity that is about 2--3 times the RMS
noise of the JVLA data.

We also spatially integrate the synthetic free-free continuum and
\hep3\ line flux densities as a function of radius to increase the
sensitivity.  The model results are shown as the dashed line in
Figure~\ref{fig:iring}.  The free-free continuum emission in our model
is a good fit to the data with a cumulative continuum flux density
that is slightly less than the value derived from our JVLA continuum
observations.  The cumulative, channel-integrated modeled \hep3\
emission is larger than the observational uncertainties at smaller
radii.  Since the contribution of the \hep3\ emission arising from the
halo may be significant, and we are unable to put stringent
constraints on the physical properties of the halo, we cannot derive
an accurate \her3\ abundance ratio limit for J320.  Nevertheless, the
synthetic \hep3\ profile in Figure~\ref{fig:model}, which is 2--3
times the RMS spectral noise, produces a limit of
$^{3}{\rm He/H} \le \nexpo{2.75}{-3}$ by number.

\begin{deluxetable}{lcccccccc}
\tabletypesize{\small}
\tablecaption{J320 NEBULA Model Parameters \label{tab:model}}
\tablehead{
\colhead{} & \colhead{$\theta_{\rm inner}$} & \colhead{$\theta_{\rm outer}$} &
\colhead{$V_{\rm exp}$} & \colhead{$T_{\rm e}$} & \colhead{$n_{\rm e}$} & 
\colhead{} & \colhead{} & \colhead{} \\
\colhead{Component} & \colhead{(arcsec)} & \colhead{(arcsec)} &
\colhead{(\kms)} & \colhead{(K)} & \colhead{(\percc)} & 
\colhead{(\hepr4)} & \colhead{(\heppr4)} & \colhead{(\hepr3)}
}
\startdata
Shell & 0.1 &  7.0 & 16.0 &  12,500 &  2100 & 0.10 & 0.00 & \nexpo{2.75}{-3} \\
Halo  & 7.0 & 25.0 & 16.0 &  12,500 &    10 & 0.10 & 0.00 & \nexpo{2.75}{-3} \\
\enddata
\tablecomments{We adopt a distance of 3.26\kpc\ to J320 based on Gaia DR2
  parallaxes \citep{chornay20}.}
\end{deluxetable}

\section{Discussion}\label{sec:discussion}

There is ample evidence that stars undergo extra mixing beyond
convection as a physical process to stir material in their interiors.
There are a variety of tracers including \he3, \li7, \cratio,
etc. observed in stars and PNe whose abundances are inconsistent with
standard stellar evolution models that only include convection.  For
low-mass stars the best candidates for this extra mixing process are
rotation-induced mixing and the thermohaline instability.  Rotation
alone is not sufficient to explain the abundance anomalies
\citep{palacios06}, but models that include both of these extra mixing
processes predict abundances at different stellar evolutionary states
that are broadly consistent with observations
\citep[e.g.,][]{charbonnel10}.  Moreover, these extra mixing processes
resolve the ``\he3\ Problem'' \citep[e.g.,][]{lagarde12b, balser18}.

One criticism of thermohaline mixing is that although we might expect
all low-mass stars to destroy their \he3\ by processing it into \he4,
there are several PNe with \her3\ abundance ratios consistent with
standard stellar yields \citep[e.g.,][]{eggleton08}.
\citet{charbonnel07b} suggested that strong magnetic fields could
inhibit thermohaline mixing.  They posited that Ap-type stars, which
have stronger magnetic fields than classical A-type stars, could
maintain their magnetic field strength as they evolve into RGB stars,
when the thermohaline instability is important.  But this may no
longer be necessary since the two most significant \hep3\ detections
in PNe, \ngc{3242} and J320, have now been shown to be incorrect.

Do all low-mass stars undergo extra mixing?  This is still an open
question.  \citet{charbonnel98a} estimated that 96\% of low-mass stars
undergo extra mixing on the RGB.  Using HIPPARCOS parallaxes, they
identified a sample of 191 stars that have passed the luminosity bump
in their evolution.  It is at this evolutionary stage when these extra
mixing mechanisms are expected to be active.  Additional processing of
material, however, could occur later in the evolution of low-mass
stars. To answer this extra-mixing question we must therefore
determine abundances in objects whose material has been fully
processed by stellar evolution: PNe.

Here we focus on \he3\ and the carbon isotopic ratio \cratio\ in PNe
to explore whether extra mixing occurs in all low-mass stars.  One
major difficulty with using PNe abundances to constrain stellar
evolution models is that the progenitor mass, also called the initial
mass, $M_{\rm i}$, is required. This is because we need to compare
abundances derived from observations in PNe with stellar evolution
models that depend strongly on the initial stellar mass.  Determining
the progenitor mass is a two step process.  First, the PN central star
mass or final mass, $M_{\rm f}$, must be determined.  This is
typically done by placing the central star on an HR diagram for
comparison with evolutionary tracks from stellar models
\citep[e.g.,][]{stanghellini93}, but there are other methods
\citep[see][]{gorny97}.  Therefore, the PN central star distance is
needed to derive the luminosity.  Second, a semi-empirical
initial-final mass relation (IFMR) is used to calculate the initial
mass given the final mass.  The IFMR is calibrated by carefully
measuring the properties of white dwarfs in open clusters where the
age and therefore cooling time can be estimated
\citep[e.g.,][]{canton18}.  Recently, \citet{marigo20} have shown that
the IFMR is not monotonic and unfortunately there is a kink in the
relationship where low-mass stars reside.

Since we need final masses to determine the initial masses we
therefore search for PNe in the literature with \her3\ or \cratio\
abundance ratios that also have an estimate of the central star mass.
We then calculate a range in $M_{\rm i}$ using the IFMR from three
different sources \citep{cummings18, canton18, marigo20}.  We must
first generate a grid of possible initial masses between
$0.83-7.22$\msun\ with an increment of 0.01\msun.  We then determine
the range of initial masses that are consistent with the PN final mass
assuming a 10\% error in $M_{\rm f}$.  The methods used to derive
these IFMRs are similar but the white dwarf samples and detailed
analyses are different \citep[for a comparison of these IFMRs
see][]{canton21}.  Our calculated ranges in $M_{\rm i}$ therefore
provide an estimate of the uncertainty and include the kink in the
IFMR discovered by \citet{marigo20}.

\subsection{\her3\ Abundance Ratio in Planetary Nebulae}\label{sec:he3}

Detecting \hep3\ in PNe is very challenging since the mass of ionized
gas in these nebulae is small, producing very weak emission line
intensities.  Typically, a detection of \hep3\ with current radio
facilities translates into an abundance ratio of
$^{3}{\rm He/H} \gsim 10^{-3}$.  This limit is either consistent with
or larger than that predicted by standard stellar models.  Therefore
\her3\ upper limits are usually not particularly useful, but there are
some exceptions (see below).  Table~\ref{tab:her3} summarizes the
properties of the three PNe with claimed \hep3\ detections.  Listed
are the \her3\ abundance ratio and estimates of the final and initial
stellar masses.  Since \ic{418} is a carbon star the progenitor mass
likely has a higher value for the lower limit than the number listed
in Table~\ref{tab:her3}; that is, $M_{\rm i} \gsim 1.5$\msun\
\citep{morisset09}.  \citet{bania21} have clearly demonstrated that
the previously claimed detection of \hep3\ in \ngc{3242} is incorrect
and therefore their limits are shown in Table~\ref{tab:her3}.  Using
the JVLA we have demonstrated here that the claimed detection of
\hep3\ in J320 is also incorrect and therefore include the limits
determined in Section~\ref{sec:results}.  Finally, we list the range
of \her3\ abundance ratios derived by \citet{guzman-ramirez16} for
\ic{418}.

\begin{figure}
  \centering
  \includegraphics[angle=0,scale=1.0]{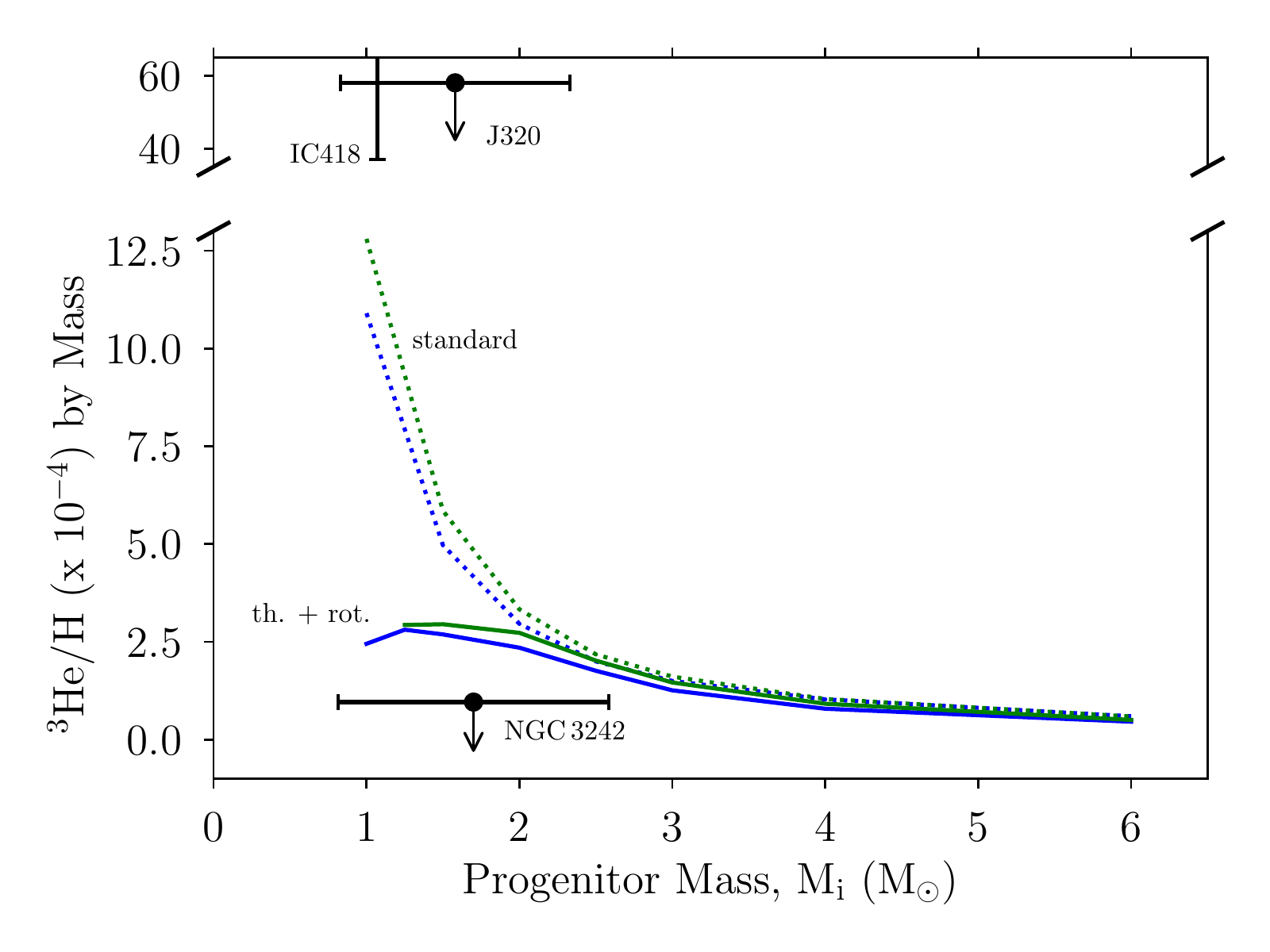}
  \caption{\her3\ abundance ratios in PNe based on observations
    (points) and stellar evolution models (curves) as a function of
    progenitor mass.  \her3\ limits are given for \ngc{3242}
    \citep{bania21} and J320 (current paper).  The lower bounds of the
    \her3\ abundance ratio are shown for \ic{418}
    \citep{guzman-ramirez16} where the progenitor mass is shifted by
    $-0.5$\msun\ for clarity.  Models are from \citet{lagarde11} where
    the dotted curves assume standard stellar evolution and the solid
    curves include thermohaline and rotation-induced mixing after the
    second dredge-up.  The blue and green curves correspond to
    metallicity ${\rm Z} = 0.004$ and 0.014, respectively.}
\label{fig:helium}
\end{figure}

We conclude that there is no longer strong evidence from PNe \hep3\
observations that {\it any} low-mass stars fail to undergo
extra-mixing.  Figure~\ref{fig:helium} plots the \her3\ abundance
ratio as a function of progenitor mass where the points correspond to
abundances derived from observations and the curves correspond to
yields from stellar evolution models.  For low-mass stars the expected
\he3\ abundances are reduced when thermohaline mixing is included.
The \her3\ upper limit for J320 is significantly larger than all
models and is therefore not very significant.  The very deep GBT
\hep3\ observations toward \ngc{3242}, however, produce a significant
\her3\ upper limit that is clearly not consistent with standard
stellar yields.  (We do not include a similar limit for the PN
\ngc{6543} since \citet{bania21} deem that this limit is not very
reliable.)  The lower range of the \ic{418} \her3\ abundance ratio
derived by \citet{guzman-ramirez16} is higher than all models.  If
this abundance is accurate then it does not support extra mixing.  As
discussed by \citet{bania21}, however, there are serious issues with
these data and this detection needs to be confirmed.

\begin{deluxetable}{llccccccc}
\tabletypesize{\small}
\tablecaption{\her3\ Abundances in Planetary Nebulae \label{tab:her3}}
\tablewidth{0pt}
\setlength{\tabcolsep}{2.0pt}
\tablehead{
\colhead{} & \colhead{} & 
\multicolumn{3}{c}{\underline{~~~~~~~~~~~~ (\her3)$\,10^{-4}$~~~~~~~~~~~ }} & 
\multicolumn{3}{c}{\underline{~~~~~$M_{\rm f}$ (\msun)~~~~~}} & 
\colhead{} \\ 
\colhead{PN} & \colhead{Alias} & 
\colhead{by number} & \colhead{by mass} & \colhead{Ref.} & 
\colhead{value} & \colhead{error} & \colhead{Ref.} & 
\colhead{$M_{\rm i}$ (\msun)} 
}
\startdata 
190.3--17.7 & J320       & {$\le 27.5$}  & {$\le 58$}    & This paper & 0.575 & \dots & G97 & $0.83-2.33$ \\ 
261.0+32.0  & \ngc{3242} & $\le 0.45$  & $\le 0.96$  & B21        & 0.615 & \dots & S20 & $0.83-2.57$ \\ 
215.2--24.2 & \ic{418}   & $17-58$     & $37-123$    & G16        & 0.573 & \dots & G97 & $0.83-2.32$ \\
\enddata
\tablecomments{Listed are the \her3\ abundance ratios, the PN central
  star mass (or final mass $M_{\rm f}$), and the PN progenitor mass (or
  initial mass $M_{\rm i}$).  We convert the \her3\ abundance ratio by
  number to mass fraction assuming a \her4\ abundance ratio by number
  of 0.1 and a metallicity of $Z = 0.0061$ for \ngc{3242} and
  $Z = 0.0122$ for J320 and \ic{418} \citep[see][]{bania21}.}
\tablerefs{B21 \citep{bania21}; G16 \citep{guzman-ramirez16}; G97
  \citep{gorny97}; S20 \citep{stanghellini20}.}
\end{deluxetable} 

\subsection{\cratio\ Abundance Ratio in Planetary Nebulae}\label{sec:cratio}

There are several different tracers used to derive the \cratio\
abundance ratio in PNe.  The brightest is the millimeter wavelength
rotational transition of CO \citep[e.g.,][]{palla00}.  There are three
problems in deriving accurate \cratio\ ratios using CO: (1) opacity
variations; (2) chemical fractionation; and (3) selective dissociation
\citep{stahl08}.  For high densities $^{12}$CO will become optically
thick and therefore the derived \cratio\ ratios will be
underestimated.  This can be mitigated by observing at least two
transitions of CO and using radiative transfer models to determine the
opacity \citep[e.g.,][]{balser02}.  Since the molecular gas in PNe is
warm ($20-50$\K) fractionation should be small \citep{ziurys20}.  But
\citet{saberi20} see variations in \cratio\ from CO due to selective
dissociation in the outflows of AGB stars and suggest that HCN is a
better tracer.  Since $^{12}$CO is more efficiently shielded than
$^{13}$CO the derived \cratio\ ratios can be overestimated when using
CO.  Nevertheless, observations of multiple molecular tracers (e.g.,
CO, HCN, CN, HCO$^{+}$, etc.) to determine \cratio\ abundances produce
results that are consistent to within the uncertainties, but there are
some exceptions \citep{ziurys20}.

A less sensitive tracer of \cratio\ abundance is the \ciii] multiplet
near $1908\,$\AA\ which has an $F = 1/2-1/2$ transition near
$1909.6\,$\AA\ that is only allowed for $^{13}$C \citep{clegg97}.
Most studies using this ultraviolet transition toward PNe, however,
only produce limits to \cratio\ \citep[see][]{rubin04}.

In Table~\ref{tab:cratio}, we list PNe with \cratio\ abundance ratios
from the literature that have estimates of the central star mass,
$M_{\rm f}$.  For PNe with \cratio\ ratios determined by different
tracers we favor HCN, CN, or HCO$^{+}$ instead of CO due to potential
issues with selective dissociation.  For PNe with multiple \cratio\
ratios based only on CO we favor those that use radiative transfer
models to derive the \cratio\ abundance.  We only include PNe that
have significant limits: $^{12}{\rm C}/^{13}{\rm C} \ge\ 5$ (see
below).  There are two independent observations of the \ciii]
multiplet transition in the well known PN NGC 3242 made using the
Hubble Space Telescope and the International Ultraviolet Explorer,
respectively.  These data yield abundance ratios of
$^{12}{\rm C}/^{13}{\rm C} \ge\ 38$ \citep{palla02} and
$^{12}{\rm C}/^{13}{\rm C} \ge\ 14$ \citep{rubin04} respectively by
number.  Inspection of the spectrum from \citet[][Figure~1]{palla02}
suggests that the quoted upper limit of
$^{12}{\rm C}/^{13}{\rm C} \ge\ 38$ is too high.  The peak-to-peak
fluctuations in the residual spectrum are about 0.2 suggesting a
$3\sigma$ limit in the $^{13}$C transition of about
$0.2 \, {\rm ergs} \, {\rm cm}^{-2} \, {\rm s}^{-1} \, {\rm
  arcsec}^{-2}$, a factor of ten larger than quoted.  We therefore
list the limit of $^{12}{\rm C}/^{13}{\rm C} \ge\ 14$ derived by
\citet{rubin04} in Table~\ref{tab:cratio}.

\begin{deluxetable}{llcccccccc}
\tabletypesize{\small}
\tablecaption{\cratio\ Abundances in Planetary Nebulae \label{tab:cratio}}
\tablewidth{0pt}
\setlength{\tabcolsep}{2.0pt}
\tablehead{
\colhead{} & \colhead{} & 
\multicolumn{4}{c}{\underline{~~~~~~~~~ \cratio ~~~~~~~~~ }} & 
\multicolumn{3}{c}{\underline{~~~~~$M_{\rm f}$ (\msun)~~~~~}} & 
\colhead{} \\ 
\colhead{PN} & \colhead{Alias} & 
\colhead{value} & \colhead{error} & \colhead{Ref.} & \colhead{Tracer} & 
\colhead{value} & \colhead{error} & \colhead{Ref.} & 
\colhead{$M_{\rm i}$ (\msun)} 
}
\startdata 
010.1+00.7 & \ngc{6537} & 2.4 & 0.30 & Z20 & HCN & 0.80 & 0.10 & M05 & $1.8-4.2$ \\ 
036.1--57.1 & \ngc{7293} & 12. & 5.4 & Z20 & HCO$^+$ & 0.71 & \dots & S20 & $1.6-3.3$ \\ 
037.7--34.5 & \ngc{7009} & $\ge 5.6$ & \dots & R04 & \ciii] & 0.60 & \dots & G97 & $0.83-2.5$ \\ 
041.8--02.9 & \ngc{6781} & 20. & 1.0 & P00 & CO & 0.82 & \dots & S20 & $2.8-4.5$ \\ 
060.8--03.6 & \ngc{6853} & $\ge 46.$ & \dots & P00 & CO & 0.71 & \dots & S20 & $1.6-3.3$ \\ 
063.1+13.9 & \ngc{6720} & 9.5 & 1.6 & B02 & CO & 0.66 & \dots & S20 & $1.3-2.9$ \\ 
084.9--03.4 & \ngc{7027} & 31. & 0.62 & B02 & CO & 0.67 & 0.030 & P00 & $1.4-3.0$ \\ 
089.8--05.1 & \ic{5117} & 14. & 1.0 & P00 & CO & 0.56 & 0.020 & P00 & $0.83-2.2$ \\ 
093.4+05.4 & \ngc{7008} & $\ge 12.$ & \dots & P00 & CO & 0.60 & \dots & G97 & $0.83-2.5$ \\ 
103.2+00.6 & M2-51 & 15. & 1.0 & P00 & CO & 0.63 & 0.090 & P00 & $0.98-2.7$ \\ 
106.5--17.6 & \ngc{7662} & $\ge 6.5$ & \dots & R04 & \ciii] & 0.68 & \dots & S20 & $1.5-3.0$ \\ 
189.8+07.7 & M1-7 & 20. & 1.8 & B02 & CO & 0.59 & \dots & S97 & $0.83-2.4$ \\ 
215.6+03.6 & \ngc{2346} & 22. & 2.7 & B02 & CO & 0.63 & 0.020 & P00 & $0.98-2.7$ \\ 
226.7+05.6 & M1-16 & 2.2 & 0.030 & B02 & CO & 0.56 & 0.020 & P00 & $0.83-2.2$ \\ 
228.8+05.3 & M1-17 & 22. & 1.0 & P00 & CO & 0.55 & 0.050 & P00 & $0.83-1.5$ \\ 
234.8+02.4 & \ngc{2440} & 1.6 & 0.50 & Z20 & HCN & 0.66 & 0.070 & M19 & $1.3-2.9$ \\ 
261.0+32.0 & \ngc{3242} & $\ge 14.$ & \dots & R04 & \ciii] & 0.61 & \dots & S20 & $0.83-2.6$ \\ 
294.6+04.7 & \ngc{3918} & $\ge 9.9$ & \dots & R04 & \ciii] & 0.62 & \dots & G97 & $0.88-2.6$ \\ 
319.6+15.7 & \ic{4066} & 20. & 3.0 & C92 & CO & 0.76 & \dots & G97 & $1.7-3.7$ \\ 
342.1+10.8 & \ngc{6072} & 12. & 3.0 & Z20 & HCN & 0.91 & \dots & G97 & $3.4-5.4$ \\ 
342.1+27.5 & Me2-1 & $\ge 6.9$ & \dots & R04 & \ciii] & 0.72 & \dots & G97 & $1.6-3.4$ \\ 
\enddata 
\tablecomments{Listed are the isotopic carbon ratio by number, \cratio, 
the PN central star mass (or final mass $M_{\rm f}$), and the  
PN progenitor mass (or initial mass $M_{\rm i}$). 
We only include significant limits (\cratio\ $\ge 5$).} 
\tablerefs{B02 \citep{balser02}; C92 \citep{cox92}; G97 \citep{gorny97}; 
M05 \citep{matsuura05}; M19 \citep{miller19}; 
P00 \citep{palla00}; R04 \citep{rubin04} 
S97 \citep{stasinska97}; S20 \citep{stanghellini20}; 
Z20 \citep{ziurys20}.} 
\end{deluxetable}

\begin{figure}
  \centering
  \includegraphics[angle=0,scale=0.8]{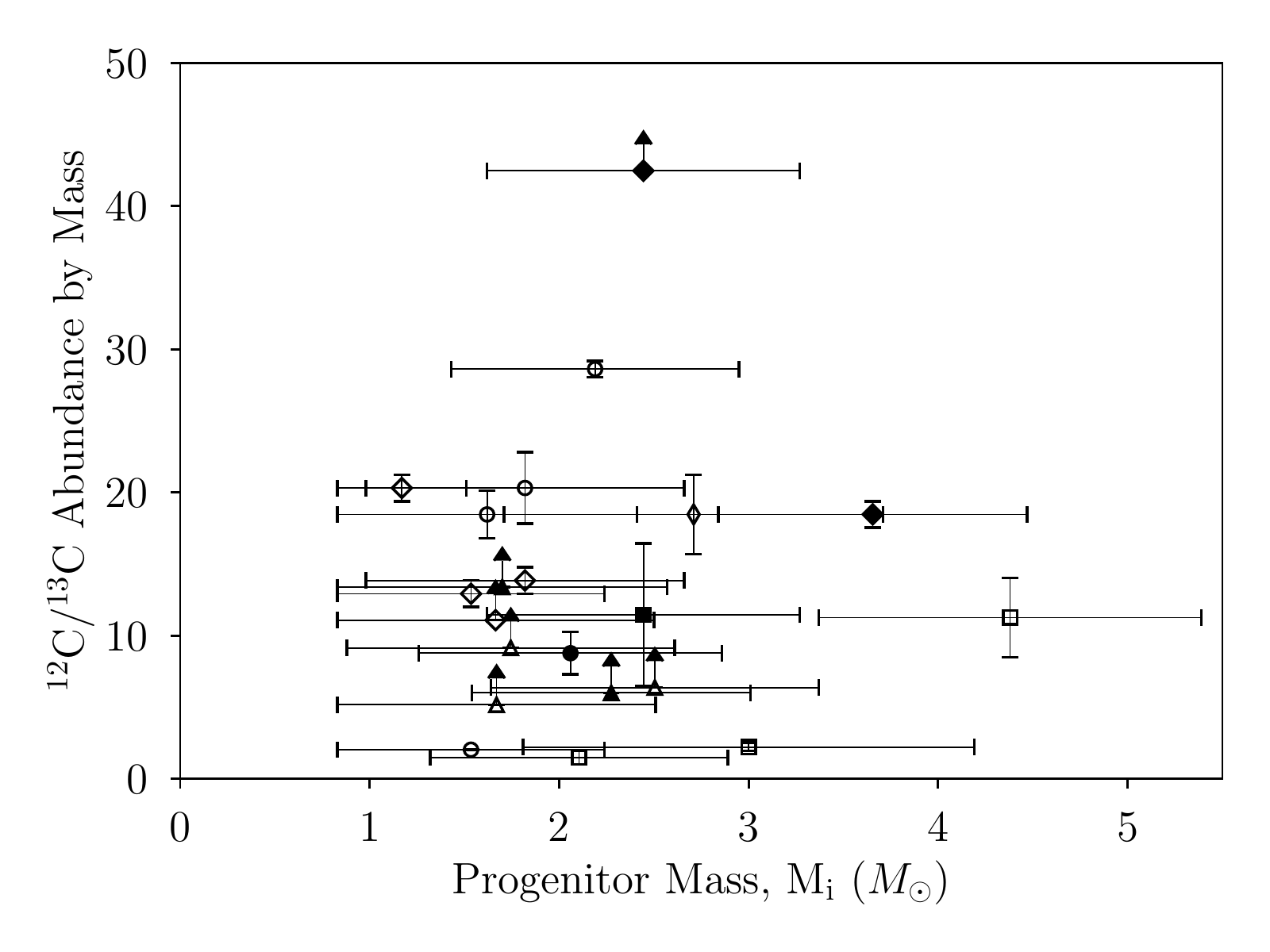}
  \includegraphics[angle=0,scale=0.8]{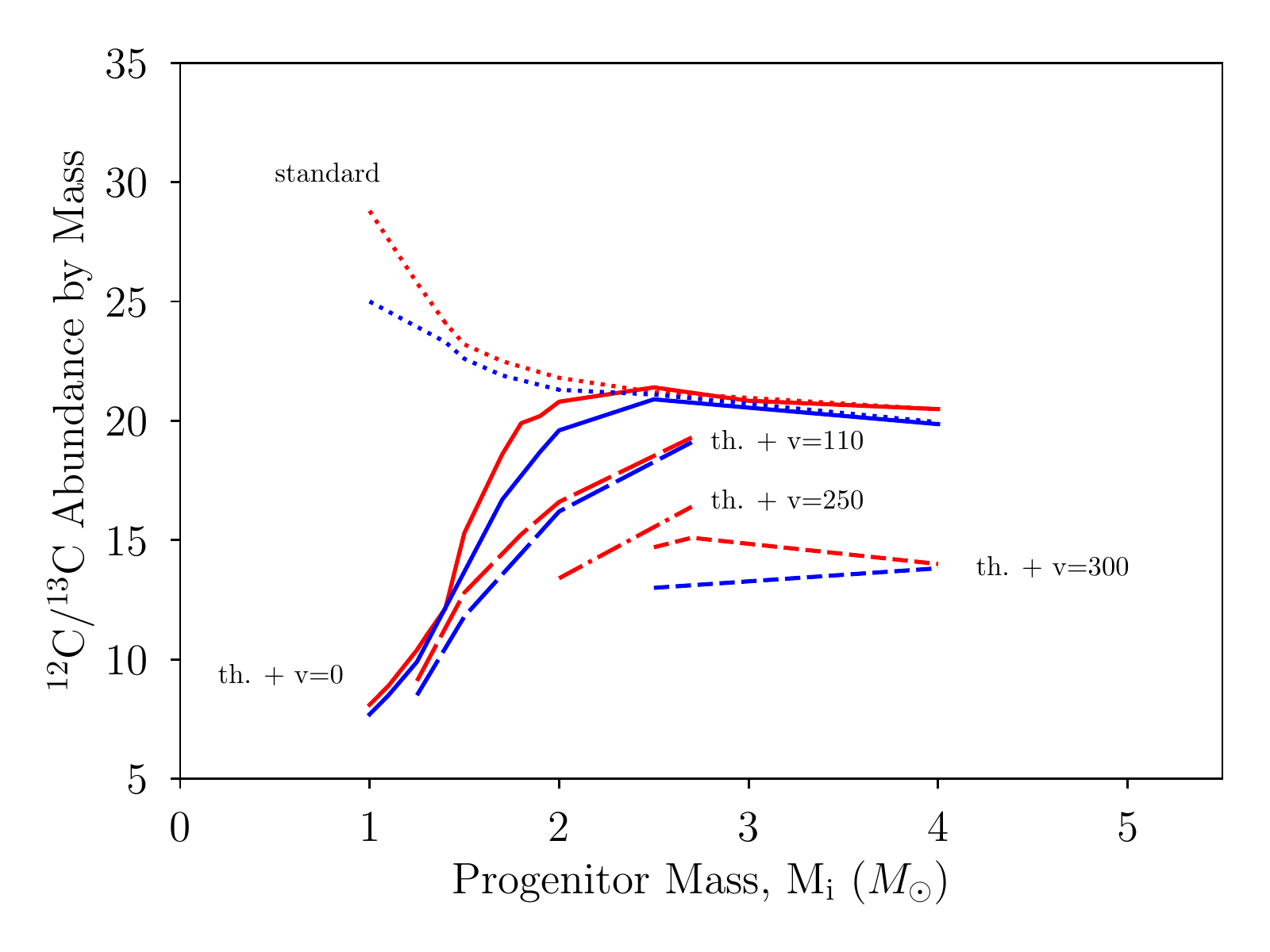}
  \caption{\cratio\ abundance ratios in PNe as a function of
    progenitor mass.  {\it Top:} \cratio\ ratios from millimeter
    molecular transitions [\citet[][thin diamond]{cox92};
    \citet[][diamond]{palla00}; \citet[][circle]{balser02}; and
    \citet[][square]{ziurys20}] and ultraviolet \ciii\ transitions
    [\citet[][triangle]{rubin04}].  Filled symbols have central
    stellar masses derived with accurate parallax-determined distances
    \citep{stanghellini20}.  {\it Bottom:} Model \cratio\ predictions
    from \citet{charbonnel10} for standard stellar evolution models
    (dotted) and models that include thermohaline mixing (solid).
    Also shown are models that include both thermohaline and
    rotation-induced mixing with various initial stellar rotation
    velocities ($v = 110$\kms, long dash; $v = 250$\kms, dash-dot; and
    $v = 300$\kms, dashed).  The red curves correspond to \cratio\
    ratios at the tip of the RGB, whereas the blue curves are \cratio\
    ratios at the end of the second dredge-up.}
\label{fig:carbon1}
\end{figure}

Figure~\ref{fig:carbon1} plots the \cratio\ abundance ratio by mass as
a function progenitor mass.  Here we convert the \cratio\ abundance
ratios listed in Table~\ref{tab:cratio} from number density to mass
fraction to compare with stellar evolution models.  This requires
multiplying the \cratio\ abundance ratios by the small factor of
12/13.  The results span a wide range of values but are concentrated
at values of $^{12}{\rm C}/^{13}{\rm C} \le 20$ and progenitor masses
between 1 and 3\msun.  As expected the progenitor masses are rather
uncertain with error bars on the order of 1\msun.  For comparison the
predictions of the \cratio\ ratio from stellar evolution models are
shown in the bottom panel.  The different curves correspond to yields
using standard models and those that include extra mixing from the
thermohaline instability and rotation.  For low-mass stars there is a
significant difference in the expected \cratio\ ratio between these
models, whereas for higher mass stars the models predict
$^{12}{\rm C}/^{13}{\rm C} \sim 20$, unless there is significant
rotation.

\begin{figure}
  \centering
  \includegraphics[angle=0,scale=1.0]{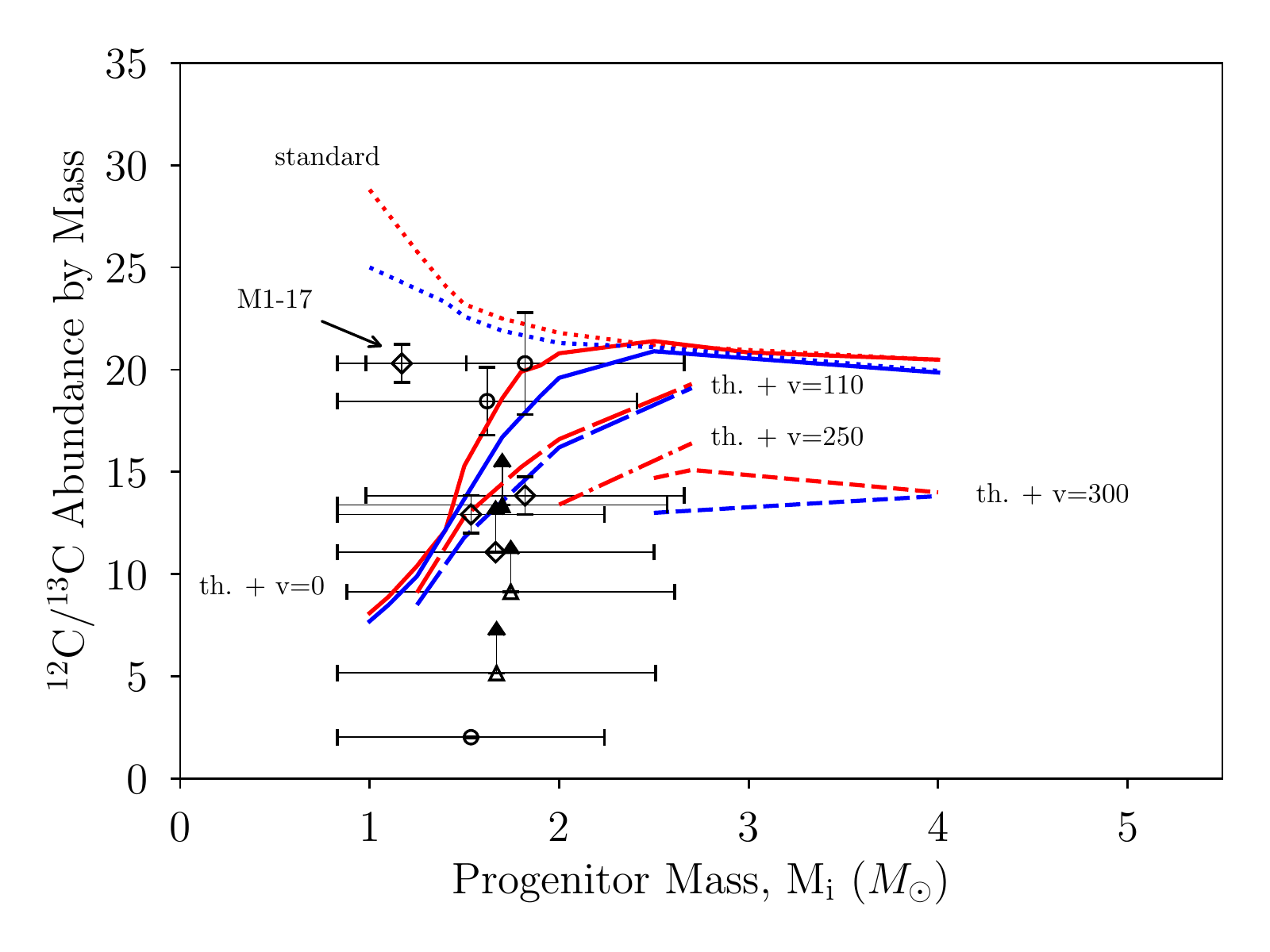}
  \caption{\cratio\ abundance ratios in PNe based on observations
    (points) and models (curves) as a function of progenitor mass.
    See Figure~\ref{fig:carbon1} for details.  We only include
    low-mass progenitor stars ($M_{\rm i} \le 2$).  All PNe with
    progenitor masses derived using parallax distances (solid points)
    require extra-mixing processes to account for their \cratio\
    abundance ratios. The PN M1-17 is labeled in the plot.}
\label{fig:carbon2}
\end{figure}

In Figure~\ref{fig:carbon2}, we combine the \cratio\ results from
observations with the predictions from models.  Since these stellar
evolution models do not include the combined effect of third
dredge-up, hot bottom burning, and thermohaline mixing for more
massive stars, we only show PNe with progenitor masses less than
2\msun.  The very low \cratio\ ratio of 2.2 for M1-16 suggests
additional mixing beyond that included in the models.  But otherwise
interpreting Figure~\ref{fig:carbon2} is difficult given the large
uncertainty in progenitor masses.  For example, a PN with
$^{12}{\rm C}/^{13}{\rm C} \sim 20$ is consistent with the standard model
for a progenitor mass of 1\msun, but the interpretation is
inconclusive if the progenitor mass is 2\msun.

There is one PN, however, that appears consistent with the standard
model: M1-17.  There are, however, some potential problems with M1-17.
First, as discussed above, \cratio\ abundance ratios can be
overestimated due to selective dissociation.  Observations of HCN, or
similar tracers, should therefore be made toward M1-17 to confirm the
high \cratio\ values.  Second, since most authors do not include an
error for the central star mass we have assumed a nominal uncertainty
of 10\%.  If we had chosen a 20\% uncertainty then the progenitor mass
range for M1-17 would be 0.83--2.5\msun\ making the results harder to
interpret.  Nevertheless, \citet{palla00} did include an error of
about 10\% for the M1-17 central star mass (see
Table~\ref{tab:cratio}).  One major source of error in determining the
central star mass is the distance.  Gaia parallaxes do exist for some
PNe central stars but alas not for M1-17.  Nevertheless, PNe
parallaxes from Gaia have been used to calibrate a Galactic PN
distance scale based on the correlation between nebular physical
radius and H$\beta$ surface brightness \citep{stanghellini20}.
Applying this new Galactic PN distance scale to M1-17 would yield more
accurate estimates for the progenitor mass.

We conclude that based on the \cratio\ abundance ratios in PNe, there
is evidence that some low-mass stars fail to undergo extra
mixing. More work is required, however, to confirm this.  In
particular, additional observations of HCN or similar tracers toward
PNe to derive \cratio\ abundance ratios are needed.  Too, unless
parallaxes are available, the new Galactic PN distance scale should be
used to derive PNe progenitor masses.

\section{Conclusions}\label{sec:conculsions}

For many decades there has been evidence that extra mixing beyond
convection must occur in low-mass stars.  Observations of \li7,
\cratio, C/O, and other tracers on the red giant branch isolate this
extra mixing to just after the star reaches the luminosity bump, when
the hydrogen burning shell reaches the chemical discontinuity created
by the maximum extent of the convective envelope during the first
dredge-up.  The two most likely candidates for this extra mixing are
rotational-induced mixing and the thermohaline instability.

Observations of \he3\ in PNe provide an important constraint to mixing
mechanisms in stars since they probe abundances in places that have
been fully processed by stellar evolution.  Standard stellar evolution
models that only include convection as a way to mix material inside
stars predict the production of significant amounts of \he3.  But GCE
models that use these standard \he3\ yields produce \her3\ abundance
ratios in the ISM that are much higher than are observed.  This
``\he3\ Problem'' can be resolved if most stars undergo extra mixing
as predicted by models that include the thermohaline instability.  Yet
there are a few PNe (\ngc{3242}, J320, and \ic{418}) with \her3\
abundance ratios consistent with the standard models indicating that
not all low-mass stars undergo extra mixing.

Recent GBT observations of \ngc{3242}, however, reveal that the
detection of \hep3\ in this PN is not real.  A mere detection of
\hep3\ is at the limit of most radio facilities and therefore each
claimed detection must be carefully scrutinized.  Moreover, the
detection of \hep3\ in \ic{418} is suspect given the lack of any
serious tests of the spectral baselines together with discrepancies in
the measured RRL parameters.

Here we observe \hep3\ at 8665.65\mhz\ in J320 made with the JVLA in
the D-configuration to confirm a previous \hep3\ detection with the
older VLA and to produce a definitive result.  Our more sensitive
observations do not detect the \hep3\ transition with an RMS noise of
58.8\microjyb.  We estimate an abundance ratio limit for J320 of
$^{3}{\rm He/H} \le \nexpo{2.75}{-3}$ by number using the radiative
transfer code NEBULA.  Based on \he3\ data there is no longer strong
evidence that some low-mass stars do not undergo extra mixing.

We also explore extra mixing by using the \cratio\ abundance ratio in
PNe.  Taking \cratio\ data from the literature we find one PN, M1-17,
that is consistent with standard stellar yields, indicating that at
least some low-mass stars do not undergo extra mixing.  The high
\cratio\ ratio of 22 in M1-17 needs to be confirmed, however, by
observations of HCN or similar tracers instead of CO which is
susceptible to selective dissociation.

\acknowledgments

We thank an anonymous referee for valuable comments that improved this
paper.  We thank Letizia Stanghellini for discussions about planetary
nebulae and the initial-final mass relation.  Emmanuel Momjian
provided technical information about the VLA and JVLA.  The National
Radio Astronomy Observatory is a facility of the National Science
Foundation operated under cooperative agreement by Associated
Universities, Inc.  This work is supported by NSF grant AST1714688 to
TMB.  This research has made use of NASA’s Astrophysics Data System
Bibliographic Services.

\vspace{5mm}
\facilities{VLA}

\software{Astropy (Astropy Collaboration et a. 2013), CASA
  \citep{mcmullin07}, Matplotlib \citep{hunter07}, NumPY \& SciPy
  \citep{vanderwalt11}, WISP \citep{wenger18}, NEBULA
  \citep{balser18}.}

\clearpage

\bibliography{ms}

\end{document}